\documentclass{aa}  
\usepackage[switch]{lineno} 
\usepackage{graphicx}
\usepackage{bm}
\usepackage{amsmath}
\usepackage{hyperref} 
\usepackage{natbib}
\usepackage{txfonts}
\usepackage{orcidlink}
\usepackage{enumitem}
\usepackage[normalem]{ulem}
\usepackage{cancel}

\begin{document} 

   \title{A wavelet-based model of magnetic turbulence over a Parker spiral field}

  \author{A.
         Larosa$^{\ast}$\orcidlink{0000-0002-7653-9147}\inst{1}
          \and
          F.
          Pucci$^{\ast}$\orcidlink{0000-0002-5272-5404}\inst{1}
          \and
          F. Malara$^{\ast}$\orcidlink{0000-0002-5554-8765}\inst{2,1,3}
          \and 
          O. Pezzi\orcidlink{0000-0002-7638-1706}\inst{1}
          \and
          S. Perri\orcidlink{0000-0002-8399-3268}\inst{2,3, 1}
          \and
          G. Zimbardo\orcidlink{0000-0002-9207-2647}\inst{2,3}
          \and 
          G. Nisticò\orcidlink{0000-0003-2566-2820}\inst{2}
          \and 
          F. Chiappetta\orcidlink{0000-0001-7221-1382}\inst{2}
          \and 
          M. Chimenti\orcidlink{0009-0004-4989-3484}\inst{2}
          \and
          L. Sorriso-Valvo\orcidlink{0000-0002-5981-7758}\inst{1,4}
          }

   \institute{Istituto per la Scienza e la Tecnologia dei Plasmi, Consiglio Nazionale delle Ricerche, Via Amendola 122/D, I-70126 Bari, Italy
         \and
         Dipartimento di Fisica, Universit\`a della Calabria, via P. Bucci, 87036 Rende (CS), Italy
         \and
         Istituto Nazionale di Astrofisica, sede di Cosenza, via P. Bucci, 87036 Rende (CS), Italy 
         \and
         Department of Electromagnetics and Plasma Physics, School of Electrical Engineering and Computer Science, KTH Royal Institute of Technology, Teknikringen 31, SE-11428 Stockholm, Sweden  
             }
   \date{Received ; accepted }

 
  \abstract
   {The propagation of Solar Energetic Particles (SEPs) through the heliosphere is strongly influenced by solar-wind magnetic turbulence, which governs particle scattering, field-line wandering, and cross-field transport. These processes shape SEP intensity profiles, anisotropies, and arrival times. Because heliospheric turbulence spans a vast range of scales, direct numerical simulations cannot capture its full spectrum. Synthetic turbulence models provide an efficient alternative by generating fields with prescribed statistical and spectral properties, enabling realistic studies of SEP transport.}
   {This work presents a numerical model able to generate synthetic magnetic turbulence in the heliosphere with the goal of studying field line and particle diffusion. The model is inspired by in situ solar wind observations, indicating that solar wind turbulence properties vary with the distance from the Sun.}
  {We extend the Cartesian wavelet-based Synthetic Turbulence Model (STM), presented in \cite{Malara2016PhRvE..94e3109M}, to the non-homogeneous expanding solar wind. This is achieved by finding a suitable change of coordinates that allows one to pass from a space in which the turbulence correlation length is homogeneous to one in which it is not. We present the analytical derivation, the numerical implementation, and testing.} 
   {We obtain an improved and computationally efficient synthetic turbulence model characterized by a background Parker spiral, a broad inertial range and tunable radial scalings for both the correlation length and the magnetic-field fluctuation amplitude, as well as a controllable level of intermittency.}
   {The model capability of reproducing solar wind turbulence up to the particle Larmor scale and to take into account solar wind radial evolution makes it a powerful tool
   to study the propagation of SEPs throughout the heliosphere.}


   \maketitle
%
\begingroup
\renewcommand{\thefootnote}{\fnsymbol{footnote}}
\footnotetext[1]{These authors contributed equally.}
\endgroup

\section{Introduction}

The heliosphere is a vast, magnetized plasma environment shaped by the continuous expansion of the solar wind from the Sun into interplanetary space. On large scales, the interplanetary magnetic field is well described by the inhomogeneous Parker spiral resulting from the combined effect of the approximately radial solar-wind and the Sun rotation \citep{parker1965dynamical}. Spacecraft observations have shown that the solar wind is pervaded by magnetic turbulence over a broad range of spatial and temporal scales, from magnetohydrodynamic (MHD) scales down to ion and electron kinetic scales \citep{bruno2013LRSP...10....2B, verscharen2019}. Heliospheric turbulence is characterized by power-law spectra, anisotropies with respect to the local magnetic field, and strong intermittency associated with coherent structures and localized enhancements of the fluctuation amplitude \citep[e.g.][]{Marsch97, Sorriso-Valvo99,Bruno03, podesta2009ApJ...698..986P, Horbury2012SSRv,bruno2013LRSP...10....2B, Chen2016JPlPh..82f5302C,Greco16, NikosIntermittency2022ApJ...934..143S, Sioulas2023ApJ...951..141S}. The heliospheric magnetic field is therefore intrinsically inhomogeneous: it contains both a large-scale, radially evolving Parker-spiral component and turbulent fluctuations whose amplitude, correlation length, and statistical properties vary with heliocentric distance \citep{Pommois2001JGR...10624965P,chenTurb2020,cuesta2022ApJ...932L..11C}.

This turbulent magnetic environment plays a central role in the dynamics of energetic particles. In the heliosphere, a particularly important population is represented by Solar Energetic Particles (SEPs), namely ions and electrons accelerated in association with solar flares, coronal mass ejections, and shock waves propagating through the corona and interplanetary medium \citep{Nitta2006ApJ...650..438N, trotta2025A&A...702A..31T}. SEP energies range from tens of keV up to the GeV range, well above the characteristic thermal energies of the bulk solar-wind plasma \citep{Reames2021}. Understanding SEP acceleration and transport is essential both from a fundamental point of view, because these particles provide information on plasma energization in turbulent and reconnecting environments, and from an applied perspective, because intense SEP events represent a major component of space-weather hazards for spacecraft, astronauts, and technological systems \citep{Reames90,Klein01,Kallenrode03,Lin05,Reames13,Desai16,Malandraki2018,Retino2022ExA....54..427R}. Once released from their acceleration sites, SEPs propagate in the turbulent heliospheric magnetic field, where they experience pitch-angle scattering, field-line wandering, perpendicular diffusion, magnetic mirroring, drift motions, and possible further energization \citep{Moraal13}. These processes determine SEP intensity profiles, anisotropies, arrival times, and longitudinal and latitudinal spreading throughout the heliosphere \citep{Ruffolo08,Laitinen19,Dalla20,Perri22}.

A central theoretical and numerical problem is therefore to describe the transport of energetic particles in turbulent, expanding, and inhomogeneous solar-wind conditions \cite[see][for a review]{whitman2023review}. A fully self-consistent treatment in which the turbulent cascade and energetic particles are evolved simultaneously over heliospheric distances, is extremely challenging. This difficulty is due to the enormous scale separation between the heliospheric system size, the turbulence correlation length, the particle Larmor radius, and kinetic plasma scales. 

Several complementary strategies have been developed to model SEP transport in the heliosphere. One class of models describes the large-scale Parker spiral and includes the effect of turbulence through prescribed transport coefficients, such as parallel and perpendicular diffusion coefficients. These coefficients may be derived from theoretical considerations, fitted to observations, or introduced phenomenologically in focused-transport or stochastic descriptions of energetic-particle propagation \citep[e.g.][]{Laitinen19,Dalla20}. Such approaches are well suited to describe SEP spreading over large heliospheric distances and to connect with spacecraft observations, but they do not explicitly resolve the particle interaction with individual turbulent structures at the gyro-scale. Another class of models is based on global or semi-global MHD descriptions of the solar wind, often including solar-wind expansion, stream interactions, and transient structures such as coronal mass ejections or shocks. Examples include heliospheric models and turbulence-transport models, such as EUHFORIA and PARADISE in which energetic particles are eventually propagated through stochastic differential equations~\citep{2018JSWSC...8A..35P, 2020JSWSC..10...57P,wijsen2022observational,wijsen2023effect,Husidic24,wijsen2025freely}, or further models in which turbulence is retained through sub-grid models and diffusion coefficients can be calculated through theoretical models such as quasi-linear theory~\citep{2017ApJS..230...21C,2025ApJ...993...87U}. These models provide an important connection between solar-wind dynamics and SEP propagation, but turbulent fluctuations responsible for particle scattering are not explicitly represented at the particle Larmor scale.

An alternative and complementary approach consists of constructing synthetic turbulent magnetic fields with controlled statistical properties and integrating energetic-particle trajectories directly in those fields. This strategy is justified by the fact that, for many SEP events, the energetic-particle density is sufficiently low that their feedback on the electromagnetic field can be ignored. This approximation makes it possible to study SEP propagation by prescribing a magnetic-field configuration and then following the trajectories of a large number of test-particles. The quality of such models depends crucially on how realistically the prescribed magnetic turbulence represents the heliospheric environment. The synthetic turbulence plus test-particle approach has been widely used in the context of cosmic-ray and SEP transport because it allows one to isolate the role of turbulence amplitude, spectral extension, anisotropy, and intermittency in determining diffusion coefficients and particle spreading \citep[e.g.][]{zimbardo2006superdiffusive, Ruffolo2006ApJ...644..971R, DeMarco2007,  puccietal2016, dundovic2020novel, mertsch2020testparticle, reichherzer2020turbulence, reichherzer2022anisotropy, pezzi2024galactic, kuhlen2025simulations}. Local synthetic-turbulence models have been implemented using different numerical techniques, including Fourier-based fields \citep{Zimbardo2000PhRvE..61.1940Z, Giacalone1999, ShalchiBook2009}, nested grids \citep{2012JCAP...07..031G, kuhlen2025simulations}, wavelet decompositions \citep{Juneja1994PhRvE..49.5179J, Cametti1998Synthetic3DTurbulence, Malara2016PhRvE..94e3109M,pucci2026cosmicray}, and multi-scale decomposition \citep{Durrive2022PhRvE.106b5307D,2024EL....14643001L, maci2024bxc, Maci2026ApJ..1005L..27M}. Such models can resolve the gyro-scale interaction between particles and magnetic fluctuations, thereby giving direct access to pitch-angle scattering, parallel and perpendicular diffusion, and the influence of coherent structures. In particular, previous studies have shown that both the extent of the turbulent inertial range and the level of intermittency can significantly influence energetic-particle diffusion \citep{puccietal2016}. 

Despite this progress, most local synthetic-turbulence models do not include the global geometry and radial evolution of the heliospheric magnetic field. Indeed, they are often formulated in Cartesian domains with a uniform background magnetic field, and are therefore not directly adapted to the Parker spiral or to the radial variation of solar-wind turbulence properties. Conversely, large-scale heliospheric transport models usually include the Parker spiral but represent turbulence and the resulting particle scattering indirectly through prescribed diffusion coefficients.  \citet{Giacalone2001JGR...10615881G} connected the flow in the photosphere to the interplanetary magnetic field turbulence, but the fluctuations are present only in the latitude-longitude plane \citep[see][for a recent application of this model]{MoradiGang2019ApJ...887..102M}. Later, \citet{Ruffolo2013ApJ...779...74R} included a slab + 2D turbulence  \citep{Bieber1996JGR...101.2511B}, but on a radial background field. Only very recently models have begun to combine a Parker-spiral background with explicit turbulent fluctuations. For example, \citet{Laitinen2023Model} introduced a model including the Parker field and a composite (slab+2D) turbulence. Despite such a model does not retain intermittency and the wave-vector space is not fully three-dimensional, the first test-particle simulations on this inhomogeneous and turbulent magnetic field have been later carried out by the same authors, investigating the angular morphology of SEPs \citep{laitinen2023solar} as well as the interplay between resonant scattering and particle drifts \citep{laitinen2025interplay}.

The purpose of the present work is to introduce a synthetic model of heliospheric magnetic turbulence that combines the large-scale Parker-spiral geometry with a broad-band, three-dimensional turbulent field whose statistical properties can be controlled. Our model builds on the Synthetic Turbulence Model (STM) introduced by \citet{Malara2016PhRvE..94e3109M}, which is based on a hierarchical wavelet-like construction of magnetic fluctuations with compact support. This approach provides an efficient way to generate turbulent fields over an extended inertial range while keeping the computational cost limited. It also allows one to control the spectral exponent and to introduce intermittency through a multiplicative-cascade prescription, following the spirit of the $p$-model \citep{Meneveau1987PhRvL..59.1424M}. The original STM was formulated in a Cartesian domain with a homogeneous background field. Here we extend it to the heliosphere by introducing a suitable curvilinear coordinate transformation that maps a homogeneous synthetic-turbulence space into a physical space characterized by a Parker-spiral background and a radially varying turbulence correlation length.

The resulting Heliospheric Synthetic Turbulence Model (HSTM) is designed to reproduce several key features of solar-wind turbulence relevant for SEP transport. First, the model includes the non-uniform Parker-spiral magnetic field as the large-scale background. Second, it allows the turbulence correlation length to vary with heliocentric distance according to observationally motivated radial scalings \citep{Pommois2001JGR...10624965P,cuesta2022ApJ...932L..11C}. Third, the root-mean-square amplitude of the turbulent fluctuations can be prescribed as a function of radial distance, making it possible to mimic different solar-wind conditions. Fourth, the model retains the possibility of tuning the level of intermittency, so that both nearly-homogeneous and strongly-intermittent fields can be generated, making a clear step forward in the modeling of synthetic magnetic fluctuations over a large Parker spiral field. Finally, the construction preserves the solenoidal nature of the magnetic field and remains computationally efficient, allowing one to cover a broad range of scales down to values comparable with energetic-particle Larmor radii.

This paper represents a substantial advancement toward a test-particle framework for SEP propagation in realistic heliospheric turbulence. In the present work, we focus on the construction, implementation, and validation of the magnetic-field model itself. The application of the HSTM to energetic-particle and magnetic-field-line diffusion will be addressed in future studies. The manuscript is organized as follows. In Section~\ref{Sec:STM} we briefly review the main properties of the Cartesian STM model. In Sections~\ref{Sec:Pre-coord} and~\ref{Sec:Curv_Coord} we introduce the heliospheric extension and derive the curvilinear coordinates that allow us to include the Parker spiral and the radial evolution of the turbulence correlation length. Section~\ref{Sec:num_implementation} describes the numerical implementation and illustrates the behavior of the generated turbulent fields, including the radial dependence of the correlation length, fluctuation amplitude, spectral properties, and intermittency. The solenoidal character of the model is discussed in Appendix~\ref{Sec:Diff_OP}. Finally, Section~\ref{Sec:Conclusion} summarizes the main results and outlines future applications to SEP transport.

\section{The Synthetic Turbulence Model: A Brief Overview}\label{Sec:STM}

The  STM \citet{Malara2016PhRvE..94e3109M} is based on the recursive subdivision of a 3D cartesian domain $
\Gamma = \left\{ (x,y,z) \,\middle|\, x \in \left[0,L_x\right] \,\vee\, y \in \left[0,L_y\right] \,\vee\, z \in \left[0,L_z\right] \right\}.
$ into smaller and smaller cells. At each subdivision the domain is divided into 8 smaller and identical cells and a new scale is introduced. This process mimic the turbulent cascade in which energy is transferred from large scales to small scales and it creates a hierarchy of cells with lengths $\{\ell_0, \, \ell_1, ...., \ell_{N_s} \}$ in a 3D lattice, where $l_0 =(L_xL_yL_z)^{1/3}$, $\ell_{m+1}=\ell_{m}/2$ (i.e. $\ell_m = \ell_0/2^m$) and $N_s$ is the number of scales included in the model. The smallest scale in the system is $\ell_{N_s}$.

The number of cells is given by

\begin{equation}
    N_{cell} = \sum_{m=0}^{N_s} 2^{3m}.
\end{equation}

The cell position within the domain is indicated by $C^{i,j,k;m}$ where $i$, $j$, $k$ indicate the position in the lattice at scale $m$.

For each cell a magnetic field fluctuation defined on the scale of the cell is defined as:
\begin{equation}\label{Eq:rot_b}
    \delta \mathbf{B}^{(i,j,k;m)}
=
a^{(i,j,k;m)} \, \nabla \times \mathbf{A}^{(i,j,k;m)} \, ,
\end{equation}
where $a^{(i,j,k;m)}$ controls the amplitude of $\delta \mathbf{B}^{(i,j,k;m)}$ and $\mathbf{A}^{(i,j,k;m)} \sim  O(l_m/l_0)$ is the normalized vector potential, whose curl magnitude is scale independent. The total fluctuating magnetic field is given by summing over all the scales and all the cells.

The components of the vector potential $\mathbf{A}^{(i,j,k;m)} (x,y,z)$ are represented by compact-support polynomial functions that vanish at the boundaries of each subdomain, together with all spatial derivatives up to fourth order. As a consequence, the resulting turbulent magnetic field $\mathbf{B}_{\rm T}$ is continuous across the entire domain, along with its spatial derivatives up to third order, guaranteeing a highly regular turbulent field. In the perspective of studying charged particle transport, this feature would allow accurate orbit integration without spurious numerical scattering. At each scale $m$ the size of subdomains is chosen to be larger than the cell itself by a factor of 2 in each direction, so that neighboring eddies overlap; this prevents the artificial vanishing of fluctuations at cell boundaries and improves statistical homogeneity. Moreover, the polynomial functional form of $\mathbf{A}^{(i,j,k;m)}$ is randomly distorted in a different manner within each cell \cite[see][ for more details]{Malara2016PhRvE..94e3109M, puccietal2016}. These local distortions are introduced to enhance the statistical homogeneity of the turbulent magnetic field $\textbf{B}_{\rm T}$ and further reduce residual lattice imprinting, while preserving the regularity of the construction.

Finally a uniform background magnetic field $\mathbf{B}_0$ is added to the fluctuating part, giving the following form for the total magnetic field:

\begin{equation}\label{Eq:fullB}
\mathbf{B}(x,y,z)
=
\mathbf{B}_0
+
\sum_{m=m_{\min}}^{N_s}
\sum_{i,j,k=1}^{2^m}
\delta \mathbf{B}^{(i,j,k;m)}(x,y,z) = \mathbf{B}_0 + \mathbf{B}_{\rm T} \, .
\end{equation}
where $m_{min}$ identifies the largest scale included in the fluctuating field $l_{max}$. The eddies in the range $\ell_{max} < \ell \le \ell_0$ are set to zero in order to improve statistical homogeneity.
The index $m_{min}= \log_2 ( \ell_0/\ell_{max})$. We choose $\ell_{max}/\ell_0 \simeq 1/10$ and $N_s = 20$, this gives a spectral width $\ell_{max}/l_{N_s} = 2^{15} \simeq 3.3\times10^4$.

The spectral exponent and the level of intermittency are controlled through $a^{(i,j,k;m)}$. Choosing $|a^{(i,j,k;m)}| \propto \ell_m^{1/3}$ gives a turbulent magnetic field with a Kolmogorov spectrum, corresponding to $\delta \mathbf{B}(\ell) \propto l^{1/3}$. 
Intermittency, i. e. the increasingly non-uniform spatial distribution of the fluctuations amplitude going towards smaller and smaller scales, is modeled as in the $p$-model by \cite{Meneveau1987PhRvL..59.1424M}, where $p$ is a fixed parameter in the range $\left[ 1/2, 1\right)$. In the $p$-model energy is transferred from larger to smaller eddies in a spatially non-uniform manner. In a similar way, in the STM at a given scale $\ell_m$, each parent eddy distributes its energy among the eight daughter eddies at scale $\ell_{m+1}$. 
Four randomly selected daughter eddies receive energy at a rate
$\epsilon_{m+1}=2p\,\epsilon_m \geq \epsilon_m$, while the remaining four receive energy at a rate $\epsilon_{m+1}=2(1-p)\,\epsilon_m \leq \epsilon_m$. 
In consequence of this multiplicative process, energy concentrates more and more at particular locations in the space when moving to smaller scales. This gives rise to localized fluctuations whose amplitude is much larger than the RMS amplitude at that scale, which is a feature related to intermittency.

When $p=1/2$, the energy transfer rate is uniform across all scales and spatial locations, $ \epsilon_{m+1}=\epsilon_m$ corresponding to a non-intermittent turbulent field. 
As $p$ increases, the disparity between transfer rates increases, leading to progressively stronger intermittency. In the present synthetic turbulence model, $p$ is treated as a free parameter that controls the level of intermittency. Consequently, the amplitude factors appearing in Eq. (\ref{Eq:rot_b}) are determined by the local energy transfer rates generated through this multiplicative cascade process as:

\begin{equation}\label{Eq:coeff}
    a^{(i,j,k;m)} = \sigma^{(i,j,k;m)} a_0 \left[ \frac{\epsilon_m^{(i,j,k;m)}}{\epsilon_0} \frac{\ell_m}{\ell_0} \right]^{1/3}.
\end{equation}

In Eq. (\ref{Eq:coeff}), $\sigma^{(i,j,k;m)}$ represents the sign, randomly chosen, of the fluctuation ($\sigma^{(i,j,k;m)} =\pm 1$), $a_0$ controls the standard deviations of the turbulent field and $\epsilon_0$ is the energy transfer rate at $l_0$.   

In the numerical implementation of the STM model the evaluation of the magnetic field is less expensive than what Eq. (\ref{Eq:fullB}) would suggest because the $\delta \mathbf{B}^{(i,j,k;m)}$ has a compact support, i. e. it is non-zero only inside a subdomain $\gamma^{i,j,k;m}$ that contains the cell $C^{i,j,k;m}$. 
$\gamma^{i,j,k;m}$ is chosen to have twice the cell size, which implies that, for any given position $\mathbf{x}=(x,y,z)$, only 8 cells at any given scale are involved in the second sum of Eq. (\ref{Eq:fullB}). The magnetic field is therefore computed as:

\begin{equation}\label{Eq:Bfast}
\mathbf{B}(\mathbf{x}) = \mathbf{B}_0 + \sum_{m=m_{\min}}^{N_{\mathrm{s}}} \sum_{\mu=1}^{8} \delta \mathbf{B}^{(\mu;\mathbf{x};m)}(\mathbf{x}).
\end{equation}
where $\delta \mathbf{B}^{(\mu;\mathbf{x};m)}$ indicates one of the eight fluctuations involved in the calculation of the magnetic field at positions $\mathbf{x}$ at the scale $m$.
Equation (\ref{Eq:Bfast}) elucidates how the use of a compact support function speeds up the calculation of the synthetic turbulent magnetic field. For example, for $N_s=14$ and $m_{min}=2$ the sum in Eq. (\ref{Eq:fullB}) contains about $10^{12}$ terms, while Eq. (\ref{Eq:Bfast}) only $28$. 

For a more detailed explanation of the STM model we refer the reader to \cite{Malara2016PhRvE..94e3109M} and \cite{puccietal2016}.

\section{Heliospheric synthetic turbulence}\label{Sec:Pre-coord}
The aim of this manuscript is to extend the STM model to generate a statistical realization of the turbulent magnetic field at any point in the inner heliosphere, from 4 solar radii out to 1 AU, while including the inhomogeneous Parker spiral as the background magnetic field.

In this respect, we address the following aspects:

\begin{enumerate}[label=(\roman*)]
    \item the presence of the non-uniform background magnetic field given by the Parker spiral;

    \item the radial variation of the correlation length $l_c$;
    
    \item the dependence of the RMS magnetic fluctuations $\delta B_{\rm RMS}$ on the radial distance $r$;
    
    \item guaranteeing that the resulting magnetic field is solenoidal. 
\end{enumerate}

We propose to achieve these goals by modifying the Cartesian STM through a suitable change of space variables. In other words, the field generated by the STM is re-expressed in terms of new curvilinear coordinates, derived in Section~\ref{Sec:Curv_Coord}, instead of Cartesian coordinates. Moreover, the amplitude of the fluctuations will be appropriately modulated as a function of the radial distance.  

Concerning point (i), we recall some important properties of the Parker spiral useful for the derivation of the curvilinear coordinates. The Parker's spiral magnetic field is given by \citep{1958ApJ...128..664P}
\begin{eqnarray}\label{Bparker}
    {\bf B}_{\rm PS} (r,\theta)= B_{{\rm PS},r}(r,\theta) \mathbf{e}_r + B_{{\rm PS},\varphi}(r,\theta) \mathbf{e}_\varphi \nonumber \\
    = B_0 \frac{r_0^2}{r^2} \mathbf{e}_r -\frac{B_0 r_0^2 \Omega}{v_{\rm sw}} \frac{\sin\theta}{r} \mathbf{e}_\varphi \, ,
\end{eqnarray}
where $(r,\theta,\varphi)$ are the spherical coordinates (with $0\le \theta \le \pi$ and $-\pi \le \varphi \le \pi$), $(\mathbf{e}_r,\mathbf{e}_\theta,\mathbf{e}_\varphi)$ are the corresponding unit vectors, $r_0$ is an arbitrary reference radial distance, $\Omega$ is the angular velocity associated with the Sun rotation, $v_{\rm sw}$ is the radial solar wind speed which is assumed to be constant, and $B_0$ is the radial magnetic field component $B_r$ at $r=r_0$.

We notice that $\mathbf{B}_{\rm PS}$ is axisymmetric (independent of $\varphi$) and has a null component along the $\mathbf{e}_\theta$ direction. Therefore, magnetic lines lie on co-axial conical surfaces, each characterized by the condition $\theta={\rm const}$. We indicate such surfaces by $\mathcal{C}_\theta$. The surface $\mathcal{C}_{\pi/2}$ coincides with the equatorial plane. On each surface $\mathcal{C}_\theta$, magnetic lines are shaped as spirals. To see this, consider the ratio $B_{{\rm PS},\varphi}/B_{{\rm PS},r}$:
\begin{equation}\label{BphisuBr}
\frac{B_{{\rm PS},\varphi}}{B_{{\rm PS},r}}=-\left(\frac{\Omega R \sin\theta}{v_{\rm sw}}\right) \left(\frac{r}{R}\right)=k \sin \theta\, \rho
\end{equation}
where $R=1$ AU, $\rho=r/R$ and $k=-\Omega R/v_{sw}$ is a constant. From now on, we will use dimensionless coordinates normalized to $R$: $\bm{\rho}=\mathbf{r}/R$.
Indicating with $d{\bm \rho}=d\rho \, {\bf e}_r + \rho \sin \theta \, d\varphi \, {\bf e}_\varphi$ a generic infinitesimal displacement on the surface $\mathcal{C}_\theta$,
the vector $d{\bm \rho}$ is tangent to a magnetic line when the following condition is satisfied:
\begin{equation}\label{cond}
\frac{\rho \sin \theta \, d\varphi}{d\rho} = \frac{B_{{\rm PS},\varphi}}{B_{{\rm PS},r}}= k\sin \theta \, \rho \, ,
\end{equation}
implying $d\varphi = kd\rho$. Integrating equation \ref{cond}, we obtain
\begin{equation}\label{spiral}
\varphi = k\rho + \chi
\end{equation}
where $\chi$ is a constant. Eq. (\ref{spiral}) indicates that the azimuthal angle $\varphi$ increases linearly with the radial distance $\rho$ when moving along a field line. Therefore, field lines are spirals drawn on the conical surfaces $\mathcal{C}_\theta$. 
Different values of the constant $\chi$ correspond to different spirals, thus $\chi$ can be used to identify a given magnetic line on a given surface $\mathcal{C}_\theta$. We also notice that the angle $\theta$ does not appear in Eq. (\ref{spiral}); therefore, magnetic lines located on different surfaces $\mathcal{C}_\theta$ but identified with the same value of the parameter $\chi$ follow the same trajectory in the $(\rho,\varphi)$ plane.

Concerning point (ii), \cite{Pommois2001JGR...10624965P} have found the following scaling law for the correlation length of the heliospheric magnetic turbulence:
\begin{equation}\label{lcvsrho}
\frac{l_c}{R}=\sigma \rho^h
\end{equation}
where $\sigma\simeq 0.033$ is a dimensionless constant and $h\simeq 1.33$ is the scaling exponent. A similar scaling ($h\simeq 1$) has been also found in more recent Parker Solar Probe data \citep{cuesta2022ApJ...932L..11C}. 
Though the spectrum of heliospheric turbulence is anisotropic \citep[see, e.g.,][for a review]{bruno2013LRSP...10....2B}, for simplicity in the present model we assume an isotropic spectrum. In particular, the correlation length $l_c$ is assumed to have the same value in all directions at a given radial distance $\rho$. The introduction of spectral anisotropy will be presented in future work.

In our model, the radial scaling of the fluctuation amplitude, point (iii), is controlled through the introduction of a scalar function dependent only on the radial distance $\rho$ that multiplies the vector potential, as discussed in Section \ref{Sec:num_implementation}. This allows the amplitude of the turbulent fluctuations of the magnetic field to scale as
$\delta B \propto\rho^\alpha$,
where the value of the index $\alpha$ can be determined by a suitable choice of a parameter in the model.

The solenoidal nature of the synthetic field, point (iv) is preserved thanks to the carefully chosen transformation of coordinates as detailed in Appendix~\ref{Sec:Diff_OP}.

\section{Curvilinear coordinates}\label{Sec:Curv_Coord}
 
The geometry of the Parker spiral magnetic lines and Eq. (\ref{lcvsrho}) will be used as a basis for the choice of a new set of curvilinear coordinates. Such coordinates are indicated by
\begin{equation}\label{curcoo}
\xi=\xi(\rho), \;\;   \zeta=\zeta(\rho,\theta), \;\; \psi=\psi(\rho,\theta,\varphi)
\end{equation}
and denoted as ``pseudoradial'' coordinate $\xi$, ``pseudolatitude'' $\zeta$ and ``pseudoazimuth'' $\psi$, respectively. In the present model, the synthetic turbulent magnetic field $\mathbf{B}_{\rm T}$ will be expressed as a function of $\xi$, $\zeta$ and $\psi$, such that the correlation length $\lambda_c$, calculated in the $\left\{ (\xi,\zeta,\psi) \right\}$ space, is uniform.
Correspondingly, we assume that the correlation length $l_c$ in the physical space is not uniform, but varies in the radial direction according to Eq. (\ref{lcvsrho}). Moreover, at any given spatial position, $l_c$ has the same value along all directions (isotropic turbulence). These conditions will be used to determine the explicit form of the curvilinear coordinates, as described below.

\subsection{Pseudoradial coordinate}
For given values of the angles $\varphi$ and $\theta$, a small increment $\delta \rho$ of the radial distance corresponds to an increment of $\xi$ given by:
\begin{equation}\label{varrel}
\delta \xi\simeq \frac{d\xi}{d\rho} \delta\rho.
\end{equation}
At any position, the correlation length $l_c$ is a small fraction of the maximum radial distance $R$. Therefore, identifying $\delta \rho$ as the normalized correlation length, namely $\delta \rho=l_c/R=\sigma \rho^h \ll 1$ and $\delta \xi$ as the corresponding correlation length in the transformed space,  $\delta \xi=\lambda_c$, from Eq. (\ref{varrel}) we derive:
\begin{equation}\label{varrel2}
\lambda_c\simeq \frac{d\xi}{d\rho} \sigma \rho^h,
\end{equation}
where $\lambda_c$ is a constant, as explained above. From this equation we obtain the condition
\begin{equation}\label{dxidrho}
\frac{d\xi}{d\rho} = \frac{\lambda_c}{\sigma \rho^h},
\end{equation}
which relates the pseudo-radial coordinate $\xi$ to the radial coordinate $\rho$.

\begin{figure}
\begin{center}
\includegraphics[width=\columnwidth]{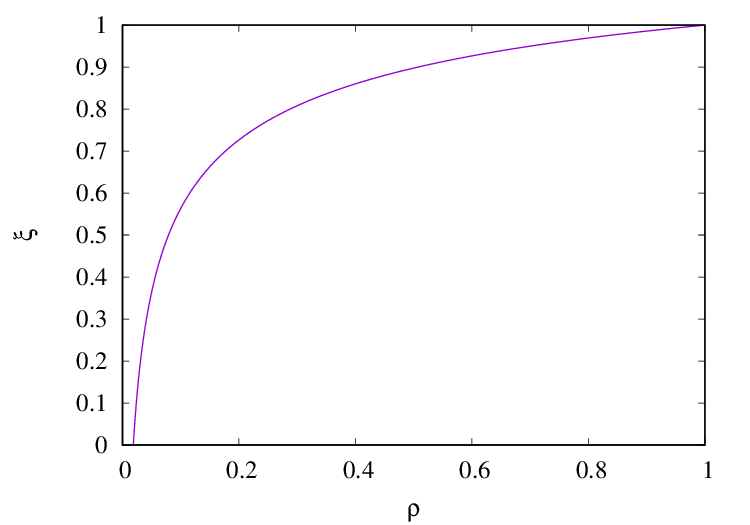}
\end{center}
\caption{The pseudoradial coordinate $\xi$ is plotted as a function of the radial coordinate $\rho$, according to Eq. (\ref{xivsrho}).}
\label{Fig:xivsr}
\end{figure}

Integrating equation (\ref{dxidrho}), we obtain:
\begin{equation}\label{xivsrho_0}
\xi=\frac{\lambda_c}{\sigma (h-1)\rho_0^{h-1}} \left[ 1 - \left(\frac{\rho_0}{\rho}\right)^{h-1}\right],
\end{equation}
where $\rho_0 = r_0/R = 4R_\odot/1$ AU $= 0.0187$ is the initial radial distance normalized to $R$. Integration constants have been chosen such that the radial distance $\rho=\rho_0$ corresponds to $\xi=0$. Moreover, we impose the condition that the distance $\rho =1$ corresponds to $\xi=1$. Using this assumption, from Eq. (\ref{xivsrho_0}) we derive the value of the correlation length in the transformed space $\lambda_c$:
\begin{equation}\label{lambdac}
\lambda_c = \frac{\sigma (h-1)\rho_0^{h-1}}{1-\rho_0^{h-1}}.
\end{equation}
Inserting this value in Eq. (\ref{xivsrho_0}) we obtain the final form of the pseudoradial coordinate:
\begin{equation}\label{xivsrho}
\xi=\xi(\rho)=\frac{1}{1-\rho_0^{h-1}} \left[ 1 - \left(\frac{\rho_0}{\rho}\right)^{h-1}\right].
\end{equation}
The function $\xi(\rho)$, plotted in Fig. \ref{Fig:xivsr}, maps the interval $\rho_0 \le \rho \le 1$ into the interval $0\le \xi \le 1$. Note that the nonlinearity of Eq. (\ref{xivsrho}) implies that equal intervals in the variable $\xi$ corresponds to unequal intervals in $\rho$.
Since the relation (\ref{xivsrho}) is monotonic, it can be inverted to obtain $\rho$ as a function of $\xi$,
\begin{equation}\label{rhovsxi}
\rho=\rho(\xi)=\frac{\rho_0}{\left[ 1 - \xi \left(1-\rho_0^{h-1}\right)\right]^{1/\left( h-1 \right)}} .
\end{equation}

The correlation length in the transformed space $\lambda_c$ is given by the expression (\ref{lambdac}). Inserting the values of $\sigma$, $h$ and $\rho_0$  we obtain $\lambda_c \simeq 0.004 \ll 1$. This implies that a number $N_{c,\rho}=1/\lambda_c \simeq 250$ of correlation lengths is contained in the radial interval $0\le \xi \le 1$ (or, equivalently, $\rho_0 \le \rho \le 1$).

\subsection{Pseudolatitude}
The pseudolatitude $\zeta$ is related to the co-latitude $\theta$. As discussed above, we aim to obtain the same correlation length in both the latitudinal and radial directions (spectral isotropy). This condition is used to define $\zeta$. We consider an infinitesimal arc length $\delta s_{\rm L} = \rho \delta \theta$ traveled in the latitudinal direction at a fixed radial distance $\rho$. 
Our goal is to obtain $\delta s_{\rm L}$ proportional to $\rho^h$, for a given variation $\delta \zeta$ of the pseudolatitude. In that case, the correlation length measured in the latitudinal direction would scale proportional to $\rho^h$, as required by the isotropy condition. 
For this purpose, we write the following relation:
\begin{equation}\label{theta1}
\theta=\frac{\pi}{2} - g(\rho) \zeta \, ,
\end{equation}
where the function $g(\rho)$ is determined by imposing the desired scaling law for $\delta s_{\rm L}$. According to Eq. (\ref{theta1}), the equatorial plane corresponds to $\zeta=0$, while the northern (southern) hemisphere corresponds to $\zeta >0$ ($\zeta <0$). Moreover, we obtain
\begin{equation}\label{deltasM}
\delta s_{\rm L} = \rho \delta \theta \simeq \rho \left( \frac{\partial \theta}{\partial \zeta} \right)_{\rho={\rm const}} d\zeta = -\rho g(\rho) d\zeta \, .
\end{equation}
Requiring $\delta s_{\rm L} \propto \rho^h$, we choose $g(\rho)=g_0\, \rho^{h-1}$, with $g_0$ a constant to be determined. From equation (\ref{theta1}), we obtain
\begin{equation}\label{theta2}
\theta=\frac{\pi}{2} - g_0\, \rho^{h-1} \zeta \, .
\end{equation}
To determine the value of the constant $g_0$, we impose that a variation $\delta \zeta = \lambda_c$ of the $\zeta$ coordinate equal to the correlation length in the transformed space corresponds to an arc length in the physical space $|\delta s_L|=l_c/R$ equal to the normalized correlation length. Using Eq. (\ref{deltasM}), we obtain the condition $l_c/R=\rho g(\rho) \lambda_c=g_0 \rho^h \lambda_c$. This gives:
\begin{equation}\label{g0}
g_0=\frac{\sigma}{\lambda_c}=\frac{1-\rho_0^{h-1}}{(h-1)\rho_0^{h-1}} \, ,
\end{equation}
where we used Eqs. (\ref{lcvsrho}) and (\ref{lambdac}). Using the values of $\sigma$ and $\lambda_c$, we find $g_0\simeq 8.25$. From Eqs. (\ref{theta2}) and (\ref{g0}) we derive the definition of the pseudolatitude in terms of coordinates $\rho$ and $\theta$:
\begin{equation}\label{zeta}
\zeta = \zeta(\rho,\theta) = \left[ \frac{(h-1) \rho_0^{h-1}}{1-\rho_0^{h-1}}\right] \frac{1}{\rho^{h-1}} \left( \frac{\pi}{2}-\theta \right)\, .
\end{equation}

The co-latitude $\theta$ can be expressed as a function of $\xi$ and $\zeta$ by inserting Eq. (\ref{rhovsxi}) into (\ref{theta2}). Using the expression (\ref{g0}), this gives:
\begin{equation}\label{thetavsxizeta}
\theta = \theta(\xi,\zeta) = \frac{\pi}{2} - \left( \frac{1-\rho_0^{h-1}}{h-1}\right) \left[ \frac{\zeta}{1-\xi \left( 1-\rho_0 ^{h-1}\right)} \right] \, .
\end{equation} 

\subsection{Pseudoazimuth}
The definition of the pseudoazimuthal coordinate $\psi$ is inspired by the spiral shape of magnetic lines. To introduce this point, we preliminarily consider the spirals defined in Eq. (\ref{spiral}), which can be rewritten in the following form:
\begin{equation}\label{chi}
\chi=\chi(\rho,\varphi)=\varphi-k\rho \, .
\end{equation}
This expression can be intepreted as a definition of a curvilinear coordinate $\chi$, related to $\rho$ and $\varphi$ through the relation (\ref{chi}).
The {\it loci} of spatial points with $\chi={\rm const}$ are the spirals, while varying $\chi$ corresponds to moving from a spiral to another in the azimuthal direction.
Using $\chi$ as a coordinate would give the advantage of taking into account the particular geometry of the problem. However, assuming that the correlation length in the azimuthal and radial direction are equal (spectral isotropy), such a choice would not give the correct dependence of the correlation length $l_c$ on the radial distance, Eq. (\ref{lcvsrho}). 
To see this, we consider an infinitesimal arc length $\delta s_{\rm A}=\rho \sin \theta \,\delta \varphi$ travelled in the azimuthal direction at constant $\rho$ on a given surface $\mathcal{C}_\theta$ ($\theta$ fixed). The quantity $\delta s_{\rm A}$ is associated with a variation $\delta \chi$. Using Eq. (\ref{spiral}) we find:
\begin{equation}\label{deltasperp}
\delta s_{\rm A}=\rho \sin \theta \, \delta \varphi|_{\rho={\rm const}} = \rho \sin \theta \, \delta \chi \, .
\end{equation}
From this equation follows that the arc length $\delta s_{\rm A}$ scales proportional to $\rho$, for a given variation $\delta \chi$ (corresponding to going from a given spiral to another on the same surface $\mathcal{C}_\theta$). In contrast, it would be desirable that $\delta s_{\rm A}$ scales proportional to $\rho^h$, similar as  $l_s$ [Eq. (\ref{lcvsrho})]. Moreover, the factor $\sin \theta$ in Eq. (\ref{deltasperp}) makes $\delta s_{\rm A}$, calculated for a given increment $\delta \chi$, to vary when moving from a surface $\mathcal{C}_\theta$ to another. This latter feature would give a correlation length in the azimuthal direction that varies in the latitudinal direction. This would be in contrast with the hypothesis that $l_c$ depends only on the radial distance.

The above considerations have guided us in the choice of the pseudoazimuth $\psi$, which is defined similar to $\chi$, but trying to obtain the correct dependence of the arc length: $\delta s_{\rm A} \propto \rho^h$. To introduce $\psi$ we modify the relation (\ref{spiral}) in the following way:
\begin{equation}\label{phi1}
\varphi=k\rho+\frac{f(\rho)}{\sin \theta} \psi \, ,
\end{equation}
where the function $f(\rho)$ will be determined by imposing the correct scaling law for $\delta s_{\rm A}$. Using Eq. (\ref{phi1}), we obtain:
\begin{equation}\label{dphi}
\delta \varphi |_{\substack{\rho={\rm const}\\ \theta={\rm const}}} \simeq \left(\frac{\partial \varphi}{\partial \psi}\right)_{\substack{\rho={\rm const}\\ \theta={\rm const}}} d\psi = \frac{f(\rho)}{\sin \theta} d \psi \, .
\end{equation}
It follows that:
\begin{equation}\label{ds2}
\delta s_{\rm A}=\rho \sin \theta \, \delta \varphi|_{\substack{\rho={\rm const}\\ \theta={\rm const}}}\simeq \rho \sin \theta \left(\frac{\partial \varphi}{\partial \psi}\right)_{\substack{\rho={\rm const}\\ \theta={\rm const}}} d\psi =\rho f(\rho)\, d\psi \, .
\end{equation}
In this way, the arc length $\delta s_{\rm A}$ calculated for a given increment $d \psi$ is independent of the co-latitude $\theta$. Moreover, to obtain $\delta s_{\rm A} \propto \rho^h$ we choose $f(\rho)=f_0 \rho^{h-1}$, where $f_0$ is a constant to be determined. Then, Eq. (\ref{phi1}) becomes
\begin{equation}\label{phi2}
\varphi=k\rho+\frac{f_0 \rho^{h-1}}{\sin \theta} \psi \, .
\end{equation}
To determine the value of the constant $f_0$ we impose that a variation $\delta \psi = \lambda_c$ of the $\psi$ coordinate equal to the correlation length in the transformed space corresponds to an arc length in the physical space $\delta s_A=l_c/R$ equal to the normalized correlation length. Using Eq. (\ref{ds2}) we obtain the condition $l_c/R=\rho f(\rho) \lambda_c=f_0 \rho^h \lambda_c$. This gives:
\begin{equation}\label{f0}
f_0=\frac{\sigma}{\lambda_c}=\frac{1-\rho_0^{h-1}}{(h-1)\rho_0^{h-1}} \, ,
\end{equation}
where we used Eqs. (\ref{lcvsrho}) and (\ref{lambdac}). Comparing Eqs. (\ref{g0}) and (\ref{f0}) we see that the two constants $g_0$ and $f_0$ are coincident.
From this relation we finally derive the definition of the pseudoazimuth $\psi$:
\begin{equation}\label{psi}
\psi=\psi(\rho,\theta,\varphi)=\left[ \frac{(h-1)\rho_0^{h-1}}{1-\rho_0^{h-1}} \right] \frac{\sin \theta}{\rho^{h-1}} \left( \varphi -k\rho \right) \, .
\end{equation}

\begin{figure}
\begin{center}
\includegraphics[width=\columnwidth]{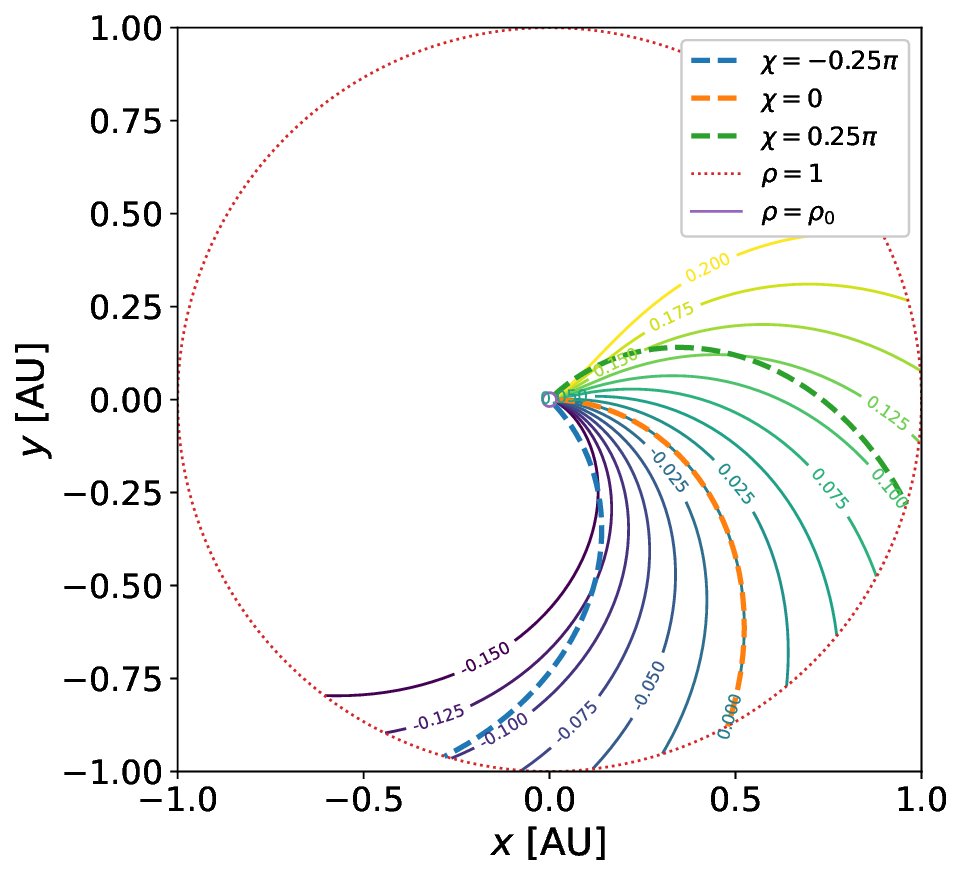}
\end{center}
\caption{Lines at $\psi={\rm const}$ are drawn in the equatorial plane, choosing $\psi$ values in the interval $\left[ -0.15, 0.2 \right]$. The long-dashed curves are Archimedean spirals and then coincide with Parker's magnetic lines in Eq.~(\ref{spiral}) for $\chi=-0.25\pi$ (blue), $\chi=0$ (orange), and $\chi=0.25\pi$ (green). The $\rho=\rho_0$ (solid purple line) and the $\rho=1$ (dotted pink line) circles are also reported.}
\label{Fig:flower}
\end{figure}

To illustrate the geometrical meaning of the pseudoazimuth $\psi$ we consider lines corresponding to $\psi={\rm const}$ in the equatorial plane $\theta=\pi/2$.
From Eq. (\ref{phi2}) we observe that the curve corresponding to $\psi=0$ is exactly an Archimedean spiral, namely, it corresponds to a magnetic line of the Parker's magnetic field. Otherwise, if $\psi\ne 0$ the corresponding curves at $\psi={\rm const}$ are deformed spirals that depart more and more from an Archimedean shape, as long as $|\psi|$ is increased.
This is visualized in Fig. \ref{Fig:flower}, where $\psi={\rm const}$ lines in the equatorial plane are plotted (continuous lines), with values of $\psi$ both positive and negative. Three Archimedean spirals satisfying Eq. (\ref{chi}) are also plotted (dashed lines). As expected, the curve, corresponding to $\psi=0$, is an Archimedeal spiral and therefore coincides with a magnetic line of the Parker field $\mathbf{B}_{\rm PS}$. All the other curves at $\psi={\rm const}$ are not exactly coincident with magnetic lines of $\mathbf{B}_{\rm PS}$. We also notice that the condition $\psi=0$ gives $\varphi=k\rho$ for any value of the co-latitude $\theta$. In other words, all the curves $\psi=0$ coincide with magnetic lines lying of the surfaces $\mathcal{C}_\theta$, regardless of the angle $\theta$.

Finally, we express the azimuth $\varphi$ in terms of the new coordinates $\xi$, $\zeta$ and $\psi$ by inserting the expressions (\ref{rhovsxi}) and (\ref{thetavsxizeta}) into Eq. (\ref{phi2}). After some algebra, this gives:
\begin{eqnarray}\label{phivsxizetapsi}
 \varphi = \varphi(\xi,\zeta,\psi) = \frac{k\rho_0}{\left[ 1-\xi \left( 1-\rho_0^{h-1}\right)\right]^{\frac{1}{h-1}}} + 
 \frac{1-\rho_0^{h-1}}{(h-1)} \nonumber \\
 \times \frac{\psi}{\left[ 1-\xi\left( 1-\rho_0^{h-1}\right)\right] \cos \left\{ \left( \displaystyle{\frac{1-\rho_0^{h-1}}{h-1}}\right) \left[ \displaystyle{\frac{\zeta}{1-\xi\left( 1-\rho_0^{h-1}\right)}}\right]\right\}}\, .
\end{eqnarray}

\begin{figure*}[htb]
\centering
\includegraphics[width=\textwidth]{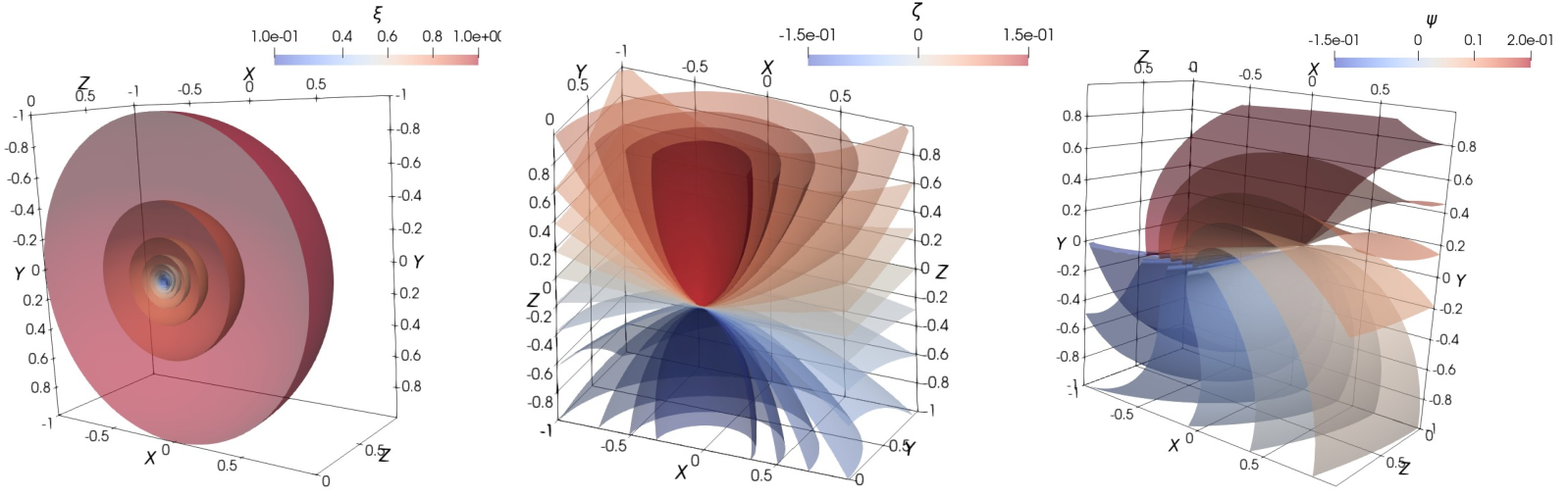}
\caption{Isocontours of the pseudo-coordinates $\xi$ (left), $\zeta$ (middle) and $\psi$ (right). The sun is at the center of the domain, and the lengths are measured in astronomical units. The isocountours of $\xi$ and $\psi$ are clipped along the equatorial plane $xy$, while the isocontours of $\zeta$ are clipped along the $xz$ plane.}
\label{Fig:zizetapsi}
\end{figure*}

In order visualize in 3D the coordinate transformation expressed by Eqs. (\ref{xivsrho}), (\ref{zeta}) and (\ref{psi}), in Fig. \ref{Fig:zizetapsi} isocontours of the pseudo-coordinates $\xi$, $\zeta$ and $\psi$ are plotted in the physical space.

It is important to note that the curvilinear coordinates described above cannot be used to represent the turbulent magnetic field throughout the spherical shell $\rho_0 \le \rho \le 1$. In fact, a jump $\Delta \varphi=2\pi$ of the azimuthal coordinate $\varphi$ is localized at the half-plane $\Pi$ defined by $\Pi =\left\{ (x,y,z) , y=0, x \le 0\right\}$. According to Eq. (\ref{psi}), this corresponds to a jump in the pseudo-space given by:
\begin{equation}\label{Deltapsi}
\Delta \psi(\xi,\zeta) = 2\pi \frac{(h-1)\rho_0^{h-1}}{1-\rho_0^{h-1}} \frac{\sin \left[\theta(\xi,\zeta)\right]}{\left[\rho(\xi)\right]^{h-1}} \; ,
\end{equation}
with $\rho(\xi)$ and $\theta(\xi,\zeta)$ given by Eqs. (\ref{rhovsxi}) and (\ref{thetavsxizeta}), respectively. Since in our model the turbulent magnetic field $\mathbf{B}_{\rm T}$ will be calculated as a function of the pseudo-coordinates, $\mathbf{B}_{\rm T}$ turns out to be discontinuous across the half-plane $\Pi$. In order to avoid this unphysical feature, we exclude a region close to $\Pi$ from the domain of interest.

\section{Implementation and properties of the model}\label{Sec:num_implementation}
The Heliospheric Synthetic Turbulence Model (HSTM) is implemented as follows:
\begin{enumerate}

       \item The method used in the model by \cite{Malara2016PhRvE..94e3109M} is applied to compute a preliminary form of the covariant components $a'_m=a'_m(q^1,q^2,q^3)$ ($m=1,2,3$) of a vector potential in the ``pseudo-space'', where $q^1=\xi$, $q^2=\zeta$ and $q^3=\psi$ indicate the curvilinear coordinates. 
       
       \item The vector potential is then corrected by multiplying by a factor $\rho^{\Delta \alpha}$ (with $\rho=\rho(\xi)$ given by Eq.(\ref{rhovsxi})), which has the role of controlling the scaling of the fluctuation amplitude with the radial distance $\rho$ (see below). Therefore, the final form of the vector potential is $a_m(q^1,q^2,q^3)=\left[\rho(q^1)\right]^{\Delta \alpha} a'_m(q^1,q^2,q^3)$, 

       \item The Cartesian components $B_{{\rm T},i}$ ($i=x,y,z$) of the turbulent magnetic field are expressed in terms of the vector potential $a_m$ through the equation \ref{cartB}. This gives:
\begin{equation}\label{cartBfr}
B_{{\rm T},i} = \frac{J_{ik}}{\mathcal{J}} \varepsilon_{klm} \frac{ \partial (\rho^{\Delta \alpha} a'_m)}{\partial q^l}\, .
\end{equation}
where $J_{ik}=\partial x_i/\partial q^k$ are the elements of the Jacobian matrix of the coordinate transformation, $\mathcal{J}={\rm det}(\left\{ J_{ik} \right\} )$, and $\varepsilon_{klm}$ is the standard Levi-Civita symbol. In Eq. (\ref{cartBfr}) summation over repeated indices is understood. A detailed derivation of Eq. (\ref{cartBfr}) is given in the Appendix.
We notice that all the quantities that appear in the RHS of Eq. (\ref{cartBfr}), namely, $a'_m$, $J_{ik}$ and $\mathcal{J}$, have explicit analytical forms as functions of the curvilinear coordinates. Therefore, for any allowed value of $q^1$, $q^2$ and $q^3$ the expression can be explicitly calculated.

 
    \item Given a point in the physical space $(x, y, z)$ in which the turbulent magnetic field value is desired, we compute the corresponding spherical coordinates $(\rho, \theta, \varphi)$,
    
    \item The corresponding pseudo-space  coordinates $(\xi, \zeta, \psi)=(q^1,q^2,q^3)$ are then computed by using Eqs. (\ref{xivsrho}), (\ref{zeta}) and (\ref{psi}), respectively.
    
    \item The Cartesian components of the turbulent field $\mathbf{B}_{\rm T}$ field are calculated at the given position by the expression (\ref{cartBfr})

    \item The Parker spiral magnetic field $\mathbf{B}_{\rm PS}$ is finally added, obtaining the total magnetic field $\mathbf{B}=\mathbf{B}_{\rm PS} + \mathbf{B}_{\rm T}$.
\end{enumerate}

\begin{figure}[htb]
\centering
\includegraphics[width=\columnwidth]{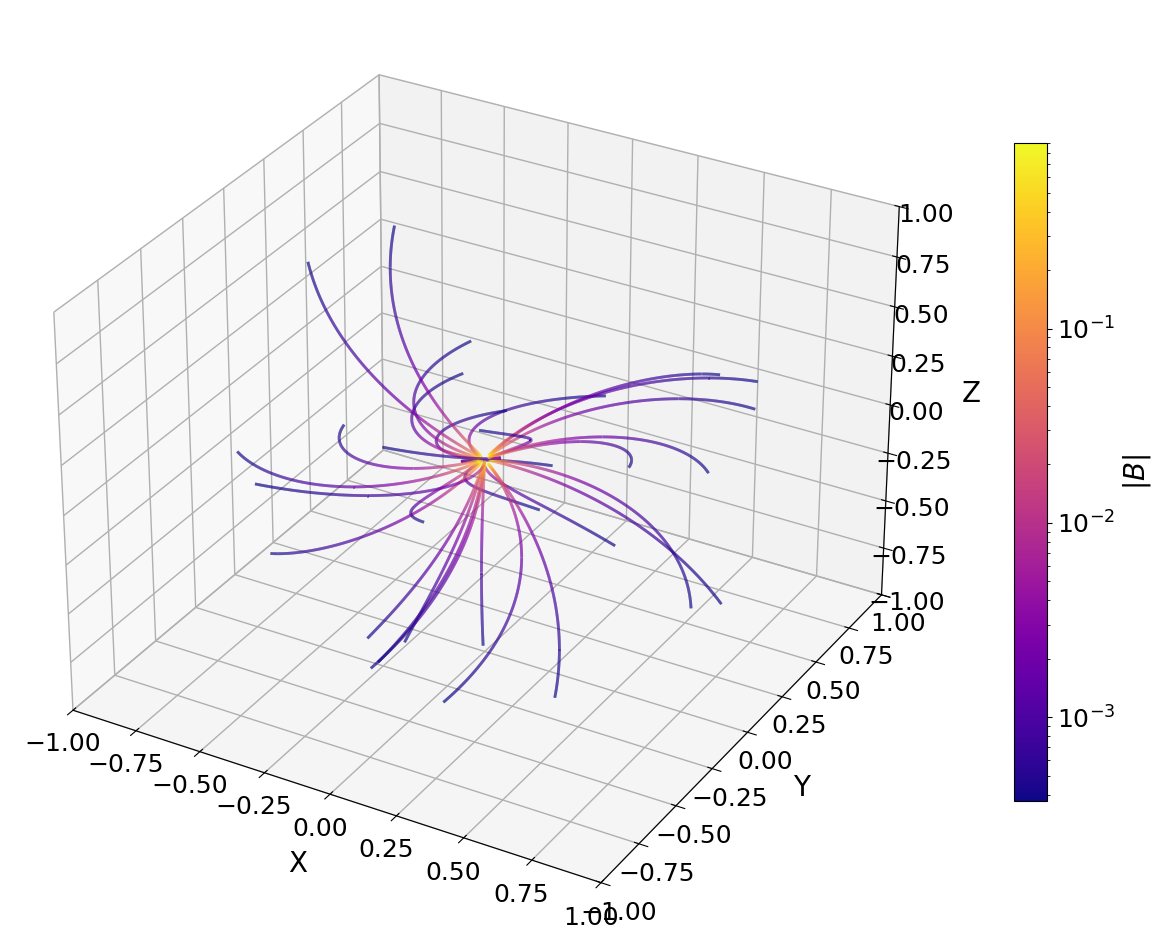}
\vspace{0.5cm}
\includegraphics[width=\columnwidth]{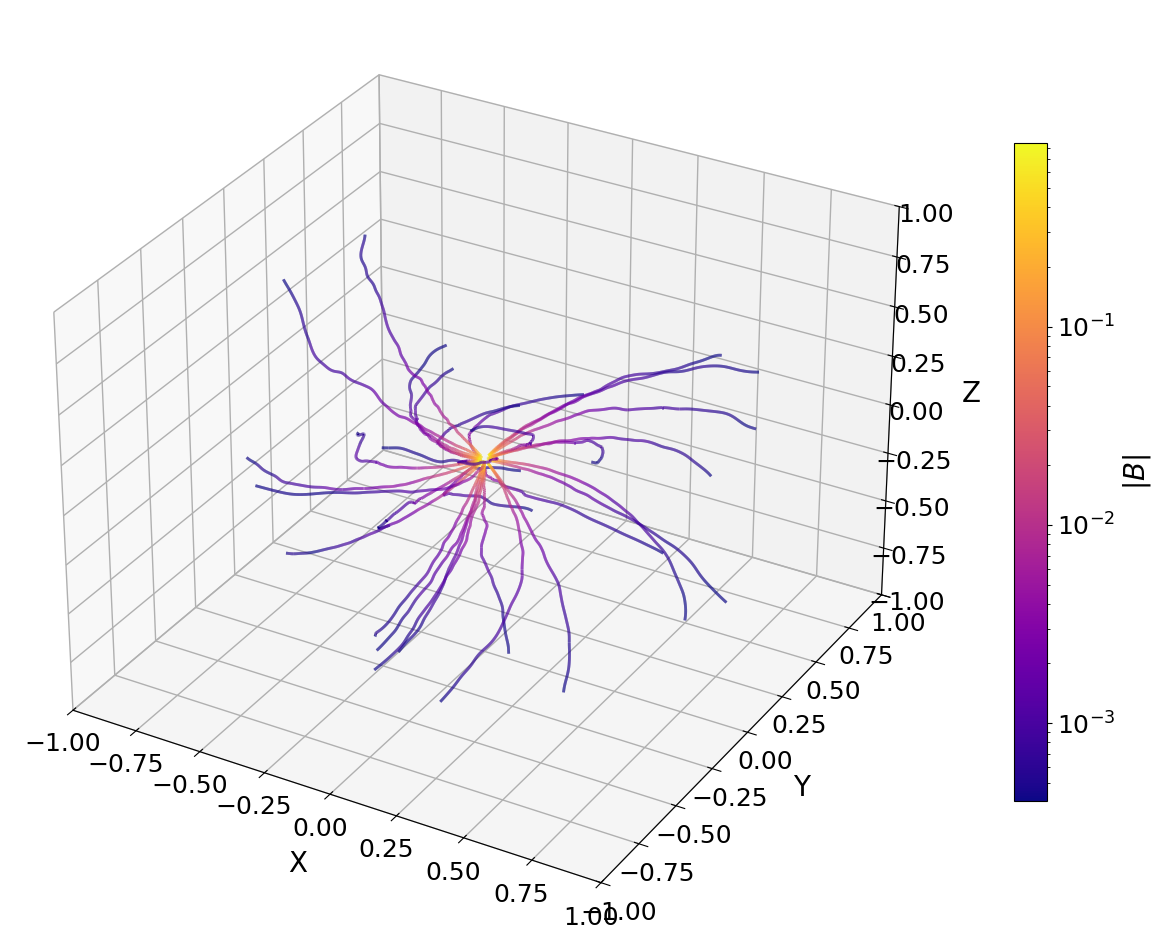}
\caption{Top: Parker spiral field lines. Bottom: Parker spiral field lines with superposed synthetic turbulence from the HSTM.}
\label{Fig:spiral_combined}
\end{figure}

The outcome of this procedure is depicted in Figure~\ref{Fig:spiral_combined}, where magnetic lines are plotted in the 3D space. In the top panel the Parker spiral field $\mathbf{B}_{\rm PS}$ lines are shown, while field lines of the total field $\mathbf{B}=\mathbf{B}_{\rm PS}+\mathbf{B}_{\rm T}$ are displayed in the lower panel. The departure from a pure Parker spiral field is clearly observable once the synthetic turbulence is added. The amplitude of the turbulent fluctuation is tunable in the HSTM, giving the possibility to reproduce realistic fluctuations in different solar wind streams.

In the following, we present and discuss the main properties of the HSTM using the parameters listed in Table~\ref{tab:parameters}. For ease of reference, Table~\ref{tab:parameters} summarizes the parameters employed to compute $\mathbf{B}$ throughout this work and provides the equation or section in which each parameter is first introduced. 

We emphasize that all the data presented here are obtained by running HSTM on a standard consumer laptop equipped with an Intel Core i7-6560U processor (2.20 GHz) and 16 GB of RAM, without the use of high-performance computing resources.

\begin{table}[h]
\centering
\begin{tabular}{lll}
\hline
\textbf{Parameter} & \textbf{Value} & \textbf{Eq. or Sec.} \\
\hline
$l_{max}$ & $l_0/2^5$ & Sec.~(\ref{Sec:STM}) \\
$l_{N_s}$ & $l_0/2^{20}$  & Sec.~(\ref{Sec:STM}) \\
$a_0$ & $0.1$ & Eq.~(\ref{Eq:coeff}) \\
$\Omega$ & $2.7\times10^{-6} \ $ rad/s & Eq.~(\ref{Bparker}) \\
$v_{sw}$ & $445 \ $ km/s  & Eq.~(\ref{Bparker}) \\
$k$ & $-0.91$ & Eq.~(\ref{BphisuBr}) \\
$h$ & $1.33$ & Eq.~(\ref{lcvsrho}) \\
$\Delta \alpha$ & $1.12$ & Eq.~(\ref{cartBfr}) \\

\hline
\end{tabular}
\caption{Model parameters and reference to the equation or section in which they are first introduced.}
\label{tab:parameters}
\end{table}

\subsection{Scaling of fluctuation amplitude}
\begin{figure}[htb]
\centering
\includegraphics[width=\columnwidth]{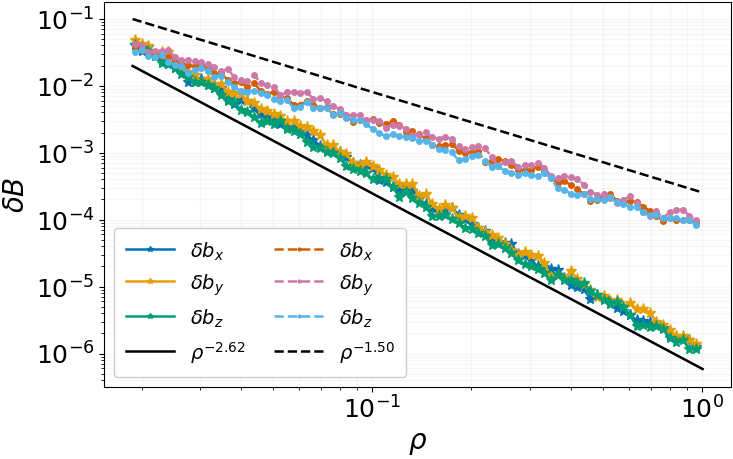}
\caption{Magnetic field fluctuations amplitudes as a function of the radial distance. The magnetic field components scale with exponent $\alpha= -2.62$ when $\Delta \alpha =0$ in Equation (\ref{cartBfr}) (solid lines). The WKB scaling (dashed lines) is obtained by setting $\Delta \alpha =1.12$.}

\label{Fig:db_vs_rho}
\end{figure}

As explained above, the radial scaling of the fluctuation amplitude is controlled by the index $\Delta \alpha$. 
In Fig.~\ref{Fig:db_vs_rho} we report how the components of the turbulent magnetic field in cartesian coordinates evolve with distance from the Sun. The $\delta B_x$, $\delta B_y$, $\delta B_z$ lines are obtained starting from $30$ different radial cuts along which the magnetic field is computed. For each cut, the large-scale Parker spiral magnetic field is subtracted from the total magnetic field to isolate the turbulent fluctuations. The absolute values of the fluctuation components are then averaged over all the radial cuts and subsequently smoothed by averaging within logarithmically spaced radial bins. 
The averaging is done solely to improve the statistics.

The results in Fig.~\ref{Fig:db_vs_rho} indicates a power-law scaling of the form $\delta B_i \propto \rho^{\alpha}$.
The radial scaling of the fluctuations is compared for the cases $\Delta \alpha=0$ and $\Delta \alpha=1.12$. The former provides the intrinsic scaling of the HSTM model as $\delta B_i \propto \rho^{-2.62}$, that is, $\alpha = \alpha^{\prime} \simeq -2.62$, where $\alpha^\prime$ is the scaling index deriving from the vector potential $a^\prime_m$. By choosing $\Delta \alpha=1.12$, we instead obtain the well-known WKB (Wentzel–Kramers–Brillouin) scaling $\alpha = \alpha_{\rm WKB}=-1.5$ \citep{parker1965dynamical}. This indicates that a generic scaling exponent $\Delta \alpha$ in Eq. (\ref{cartBfr}) provides an observed scaling in magnetic field fluctuations according to the simple formula:
$\alpha = \alpha^\prime+\Delta \alpha$. Hereafter, we adopt $\Delta \alpha=1.12$ that gives $\alpha=\alpha_{\rm WKB}=-1.5$.


\subsection{Correlation length}
We compute the autocorrelation function as is customarily done through

\begin{equation}\label{Eq:AutocorrData}
C(\Delta \ell)=
\frac{1}{L-\Delta \ell}
\frac{\int_{s_0}^{s_0+L-\Delta \ell}
\mathbf{B}(s+\Delta \ell) \cdot \mathbf{B}(s)
}{\int_{s_0}^{s_0+L-\Delta \ell}
|\mathbf{B}(s)|^2
}
\,ds ,
\end{equation}
where $s$ is a curvilinear coordinate that can be evaluated in principle along any direction, $L$ represent the length of the curve, $s_0$ the starting point of the calculation, and $\Delta \ell$ the value of the increment along the curvilinear coordinate.

\begin{figure}[htb]
\centering

\includegraphics[width=\columnwidth]{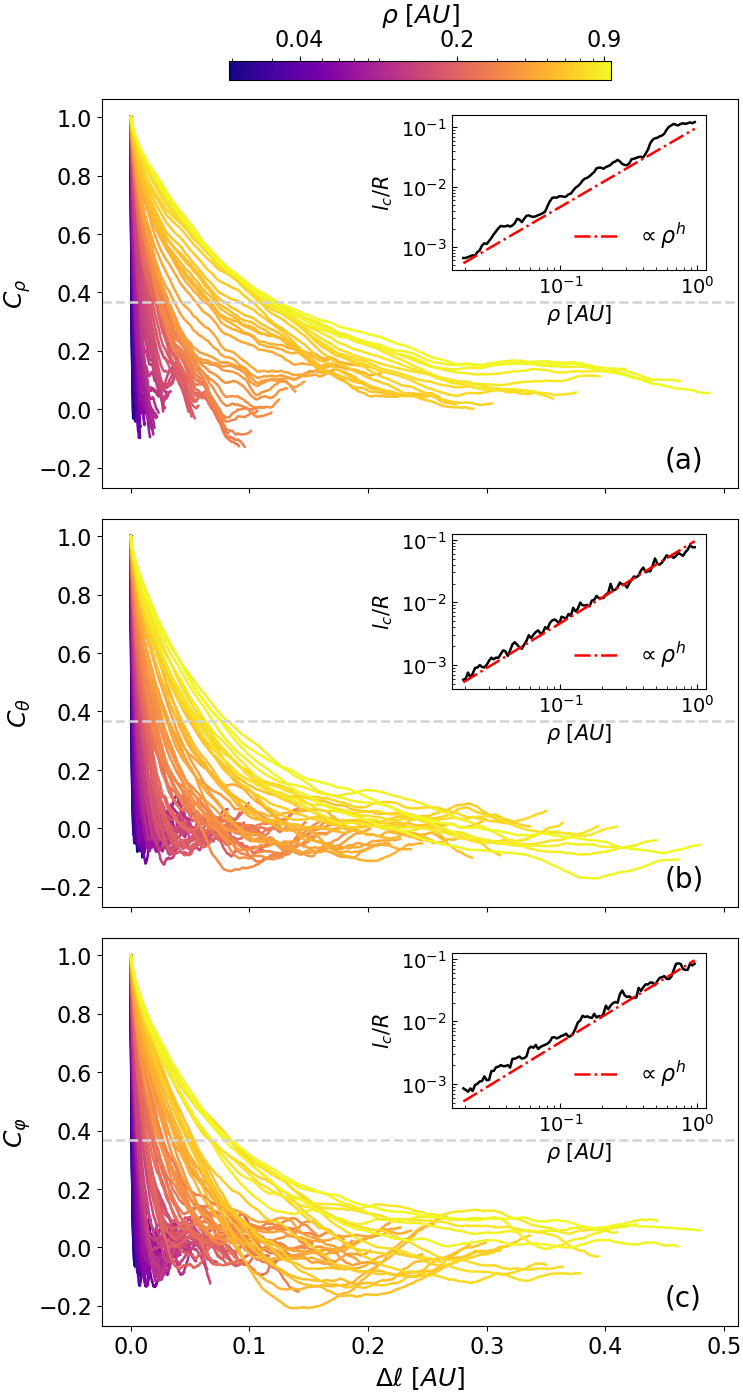}

\caption{
Autocorrelation functions computed with lags along different directions:
(a) radial direction,
(b) arcs along the $\theta$ direction,
(c) arcs along the $\varphi$ direction. The light gray lines correspond to a $1/\mathrm{e}$ value. The corresponding lag for each distance (i. e. the correlation length in the physical space) is plotted in the insets with respect to radial distance. }

\label{Fig:autocorrelation_all_directions}
\end{figure}

Figure~\ref{Fig:autocorrelation_all_directions}, reports the correlation length computed by varying $s$ along the radial direction (a), along the latitudinal direction (b) and along the azimuthal direction (c) in the physical space. These choices correspond respectively to increments of the form $\Delta \ell = \Delta \rho $ (a), $\Delta \ell = \rho \Delta \theta$ (b), and $\Delta \ell = \rho \sin{\theta} \Delta \varphi$ (c).

The smoothness of the curves is due to an average procedure. When the increments are computed radially we average over different values of $\theta$ and $\varphi$ (4 values each in Figure~\ref{Fig:autocorrelation_all_directions} (a)), when they are computed along $\theta$ we keep $\rho$ fixed, but we average each curve for 20 different values of $\varphi$ (b) and vice versa for increments along $\varphi$ (c).
In the latter case, extra care is needed since there is a $\sin \theta$ dependence in $\Delta \ell = \rho \sin{\theta} \Delta \varphi$ . To avoid averaging different $\Delta \ell$ values (while moving along the azimuthal direction at different $\theta$ values) we choose to use $\Delta \varphi  \propto 1/\sin{\theta}$. 
The averaging is intended solely to improve the statistics. The horizontal black line corresponds to 
$C(l_c/R) = 1/\mathrm{e}$.

Figure~\ref{Fig:autocorrelation_all_directions} illustrates how the 
autocorrelation function becomes broader with increasing radial distance.
The radial scaling of the latter (insets in the figure), as prescribed by Eq. (\ref{lcvsrho}), has the form $l_{c,\rho} \, , l_{c,\theta} \, , l_{c,\varphi} \propto \rho^h$, independently of the direction along which the increments are computed. 
Therefore, the desired scaling law of the correlation length as a function of the radial distance is correctly reproduced by the model, as well as the isotropy. If desired, a different scaling can be obtained by simply choosing a different value of the free parameter $h$.  

Recently, the radial evolution of the correlation length of the solar wind turbulence has been revisited by including novel PSP data \citep{cuesta2022ApJ...932L..11C}. While anisotropy relative to the magnetic field direction is present, the ratio between the parallel and perpendicular correlation length does not depart significantly from unity within 1 AU. Therefore, the isotropic correlation length in the HSTM model is a valid approximation. 

\subsection{Power spectral density}
\begin{figure}[htb]

\centering
\includegraphics[width=\columnwidth]{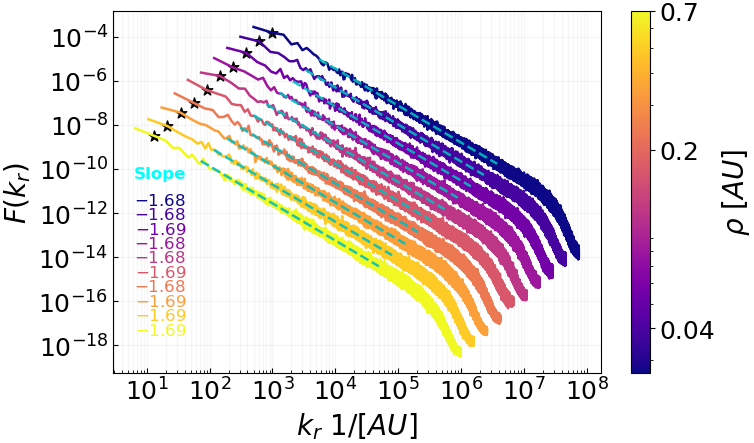}
\caption{Energy spectral density of $\mathbf{B}_{\rm T}$ calculated at different radial distances. Fitting power-laws are indicated by dashed lines, along with the corresponding spectral indexes. Black stars indicate the correlation length value $R/l_c$.
}
\label{Fig:trace_psd}

\end{figure}

We compute the power spectrum by sampling the heliospheric synthetic turbulence model in equispaced points located along radially directed lines in the physical space. The sample size varies with the radial distance from the sun to be equal to a few ``local'' correlation lengths, and the grid step is chosen to resolve the variation at the smallest-scale wavelet. The inertial range of the STM realization we used is approximately $4$ orders of magnitude wide. The size of the latter is a free parameter of the model. We recall that, as in \citet{Malara2016PhRvE..94e3109M}, the computational cost increases linearly with the number of scales, enabling the reproduction of inertial ranges of 5 orders of magnitude wide \citep[see, e.g.,][]{puccietal2016}, as in the solar wind, while keeping the computational cost low. 
To generate one of the 10 spectra plotted in Figure~\ref{Fig:trace_psd}, we compute the field in $4.8\times10^6$ points. The latter operation takes approximately 3 minutes
on a laptop equipped with the previously mentioned hardware specifications.

The evolution with radial distance of the energy spectral density (square of the Fourier coefficients) is shown in Figure~\ref{Fig:trace_psd}, where $k_r$ indicates the radial wave vector of fluctuations. It can be observed that the fluctuation power is higher at lower heliocentric distances, in accordance with Fig. \ref{Fig:db_vs_rho} and with solar wind observations.

The shift of the curves to smaller $k_r$ with increasing radial distance is due to the corresponding variation of the local correlation length $l_c$ (the value $R/l_c$ is indicated by stars in the figure). As $l_c$ increases with increasing radial distance, the spectrum is shifted toward larger spatial scales. This behaviour is consistent to what observed in the solar wind turbulence \citep{bruno2013LRSP...10....2B}. 

The spectral exponent is determined by the $1/3$ exponent in Eq. (\ref{Eq:coeff}) which gives a $-5/3$ Kolmogorov spectrum. Figure~\ref{Fig:trace_psd} shows that this feature is consistently reproduced in the fluctuation spectra at all radial distances. Note that the spectra in Fig.~\ref{Fig:trace_psd} are computed only for $\mathbf{B}_{\rm T}$, if computed on $\mathbf{B}=\mathbf{B}_{\rm PS} + \mathbf{B}_{\rm T}$ the radial trends are not affected, although the spectral exponent slightly steepens to $\sim -1.8$ at the closest distance and to $\sim -1.76$ at the furthest distance in Fig.~\ref{Fig:trace_psd} (yellow line).   

Concerning the spectral index in the solar wind turbulence, early measurements form Parker Solar Probe in the inner Heliosphere \citep{chenTurb2020} have shown that the magnetic field spectral index evolves with the radial distance, evolving from $-3/2$ closer in to $-5/3$ further out, and that the controlling parameter is the cross-helicity \citep{JackCross2023}. On the other hand, by following a corotating stream coming from the same coronal source, the radial evolution of the magnetic field power spectral slope has been observed to change from $-5/3$ close to the Sun to $-3/2$ at larger distances \citep{Perrone22}, so that a unique trend cannot be deduced. Furthermore, many authors realized that, within the inertial range itself, the spectral index switches from $-3/2$ at larger scales to $-5/3$ at smaller scales. This happens at a given break scale that depend on the radial distance \citep{Nikos2025ApJ...993..142S, Wu2025A&A...697A.187W, Mondal2025ApJ...982..199M}. 
The physical origin of this phenomenon is still unknown \citep{Nikos2025ApJ...993..142S}, therefore it is not implemented in the HSTM model at this time. 
The radial evolution of the spectral index will be implemented in a successive version of the code.

Note that, at present, the HSTM only models the inertial range portion of the solar wind spectrum. The decrease in spectral power observed at large $k$ arises solely from the absence of energy at those scales in the simulation shown in Fig.~\ref{Fig:trace_psd}. This cutoff depends on the number of scales in the wavelet hierarchy, the drop in the spectral energy happens at the typical wavenumber associated to smallest wavelet scale.

\subsection{Intermittency}
In the HSTM model, as in the STM model \citep{Malara2016PhRvE..94e3109M}, the intermittency of the fluctuations is implemented through a $p$-model \citep{Meneveau1987PhRvL..59.1424M}, where $p$ is a free parameter in the interval $1/2 \le p <1$. 
Values of $p$ larger than $p=1/2$ redistribute the energy unequally between different eddies in the cascade process, giving rise to the intermittent behavior of the signals generated through the STM \citep[see][for a more detailed discussion.]{Malara2016PhRvE..94e3109M}. Larger values of $p$ corresponds to a higher level of intermittency.

We evaluate the intermittency of magnetic field fluctuations generated by HSTM, as well as the intermittency evolution with the distance from the Sun, by computing the scale dependent Kurtosis \citep{Frisch1995}:

\begin{equation}\label{scaleKurt}
\mathcal{K}_i (\Delta \ell)=
\frac{
\displaystyle
\frac{1}{L-\Delta \ell}
\int_{s_0}^{\,s_0+L-\Delta \ell}
\left|
B_i(s+\Delta \ell)-B_i(s)
\right|^4
\,ds
}{
\left[
\displaystyle
\frac{1}{L-\Delta \ell}
\int_{s_0}^{\,s_0+L-\Delta \ell}
\left|
B_i(s+\Delta \ell)-B_i(s)
\right|^2
\,ds
\right]^2
},
\end{equation}
where $s_0$, $L$ and $s$ have the same meaning of Eq. (\ref{Eq:AutocorrData}). The index $i$ indicates the different components of the field in either cartesian $(x, y, z)$ or spherical $(\rho, \theta, \varphi)$ coordinates.

The Kurtosis can be computed across different direction, for different components and for different values of the parameter $p$. Unlike the non-intermittent case generated by the STM \citep{Malara2016PhRvE..94e3109M}, in the HSTM, the Kurtosis slightly departs from the Gaussian value $\mathcal{K} = 3$ even for $p=1/2$, showing the presence of weak residual intermittency. We investigated in depth this feature, which, however, becomes milder when the kurtosis is computed over many different realizations of the fields. We believe that this feature arises from the change of coordinates necessary to introduce a radial dependence in the correlation length. A non-Gaussian fluctuation seems to be the price of representing radially evolving turbulence in our implementation. However, the level of kurtosis introduced by our procedure remains close to $\mathcal{K} = 3$ and much smaller than what observed in the solar wind and in our model for $p\neq\frac{1}{2}$.

\begin{figure}[htb]
\centering

\includegraphics[width=\columnwidth]{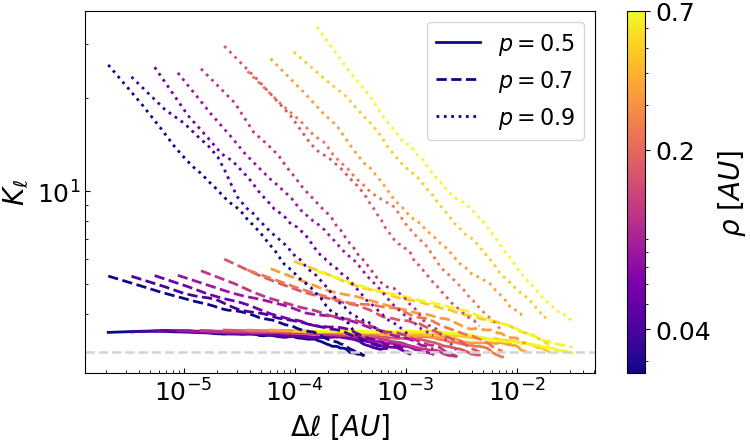}

\caption{
Kurtosis computed with lags along the Parker Spiral direction color coded with respect to radial distance. Solid lines are from the intermittency-free run $p=0.5$, dashed lines correspond to $p=0.7$ and solid lines to $p=0.9$.
The lines are averages over 40 Parker spiral arms. The dashed gray line indicate the value $K_\ell=3$.
}
\label{Fig:kurtosis_along_PS}
\end{figure}

In Figure~\ref{Fig:kurtosis_along_PS} we show the radial evolution of the kurtosis (computed for each magnetic field component and averaged) along different Parker spiral arms. The choice of computing the increments along the spiral is motivated by the fact that these are the fluctuations that a solar energetic particle would experience moving out from the Sun along the Parker spiral.
We can observe that, for $p=\frac{1}{2}$ (solid lines), the Kurtosis weakly increases towards small scales, departing from the expected non-intermittent value $\mathcal{K}=3$, but reaching at most a value of $3.6$. For moderately ($p=0.7$, dashed lines) and  strongly intermittent turbulence ($p=0.9$, dotted lines), the Kurtosis increases at small scales following the same power-law trend at all distances. The $\mathcal{K_\ell}$ values are similar to those observed in the solar wind \citep{Telloni_PSP_SO_2021ApJ...912L..21T, NikosIntermittency2022ApJ...934..143S}.

The shift in the curve positions with radial distance has the same origin as that observed in the power spectra. The lags are defined as submultiples of the local correlation length, which increases with radial distance. Consequently, the curves shift toward larger length scales. This trend is opposite to that seen for the power spectrum because the horizontal axis is a length scale instead of a wavevector.

The behavior of the kurtosis is similar when the increments are computed along the $\theta$ or $\varphi$ directions, while along the radial direction, for the non-intermittent case, a slightly more pronounced deviation from $3$ is observed (with $\mathcal{K_\ell}\lesssim5$). The latter is possibly due to the fact that the correlation length varies with the radial distance. Therefore, turbulence cannot be considered homogeneous while moving in the radial direction.

\section{Discussions and Conclusions}\label{Sec:Conclusion}

We present the Heliospheric Synthetic Turbulence Model (HSTM), a wavelet-based model that generates synthetic turbulent magnetic fields superimposed on a large-scale Parker spiral magnetic field.

HSTM is inspired by an already existing Cartesian model of synthetic turbulence \citep{Malara2016PhRvE..94e3109M}, which we have extended to the case of the expanding solar wind.
The inhomogeneous background is accounted for through the analytical derivation of a suitable change of coordinates (see Section~\ref{Sec:Curv_Coord}). The latter introduces a radial dependence of the turbulent properties on the distance from the sun while preserving the divergence-free condition for the magnetic field (see Appendix~\ref{Sec:Diff_OP}). 

As we show in the paper, the model has the following properties:
\begin{enumerate}[label=(\roman*)]

\item the correlation length scales with the distance from the sun according to a power-law whose exponent is a free parameter of the model (Fig~\ref{Fig:autocorrelation_all_directions});

\item the magnetic spectra present a broad inertial range, up to the realistic solar wind spectral extension, which follows a power-law whose exponent is a free parameter of the model (Figure~\ref{Fig:trace_psd});

\item the model can account for the presence of intermittency by generating fluctuations whose kurtosis is scale-dependent and increases at small scales (Figure~\ref{Fig:kurtosis_along_PS});

\item the amplitude of magnetic fluctuations varies as a function of the radial distance from the sun according to a power-law scaling that is tunable (Figure~\ref{Fig:db_vs_rho});

\end{enumerate}

Our model's main strength lies in its ability to reproduce a spectral range that extends to scales comparable to the solar wind, thereby including magnetic fluctuation power at the Larmor scale. The latter is enabled by the wavelet method, inherited from the old Cartesian model, which allows for an enlarged spectral extension at a reduced computational cost compared with Fourier-based methods. 

The turbulent field generated by HSTM populates the full spectral space in wave-vector space, a feature not present in reduced approaches such as 2D and 2D plus slab. Moreover, the model allows for non-Gaussian fluctuations and a controllable scale-dependent kurtosis.

The model, however, presents the following limitations:
\begin{itemize}
    \item the local correlation length is isotropic, meaning that at a given point in space the largest turbulent eddy has the same extension in all directions; 
    
    \item the spectra are also isotropic in wavenumber and polarization of the fluctuations;
    
    \item even though small, a residual non-Gaussian kurtosis is present in the case in which the intermittency is turned off, which is probably a result of the change of coordinates. 

\end{itemize}
Extension of the model to address these limitations is left for future work.
Despite these limitations, our model is a suitable tool for studying the transport of energetic particles throughout the heliosphere.  

The HSTM model was tailored to reproduce solar wind properties, however it is general enough to study particle propagation in other astrophysical contexts where turbulence properties are not homogeneous. For instance, the model can be applied to other planetary systems for which the large-scale magnetic field structure is inferred from numerical simulations, which cannot model the small-scale magnetic field \citep{vidotto2015on,farrish2019stellar}. 

The application of the HSTM to the study of SEP propagation in the heliosphere is the next natural step, along with the exploration of its applicability to other astrophysical environments.

\section*{Acknowledgements}
AL, FP, FM, SP, GZ, GN, FC, MC, and LSV acknowledge the project ‘Data-based predictions of solar energetic particle arrival to the Earth: ensuring space data and technology integrity from hazardous solar activity events’ (CUP H53D23011020001) ‘Finanziato dall’Unione europea – Next Generation EU’ PIANO NAZIONALE DI RIPRESA E RESILIENZA (PNRR) Missione 4 “Istruzione e Ricerca” - Componente C2 Investimento 1.1, ‘Fondo per il Programma Nazionale di Ricerca e Progetti di Rilevante Interesse Nazionale (PRIN)’ Settore PE09.
SP was supported by the International Space Science Institute (ISSI) in Bern, through ISSI International Team project \#24-608 ``Energetic Particle Transport in Space Plasma Turbulence".
LSV was supported by the Swedish Research Council (VR) Research Grant N. 2022-03352, by the Swedish National Space Agency (SNSA) Research Grant N. 2025-00229, by the Agenzia Spaziale Italiana (ASI) project ``Radial Evolution of large- and KInetic-scale Processes in the Expanding Solar wind (REKIPES)'' C83C25000900005, and by the International Space Science Institute (ISSI) in Bern through ISSI International Team project \#23-591 ``Evolution of Turbulence in the Expanding Solar Wind''. 
OP are supported by the FIS2 Starting Grant FIS-2023-00246 “PhAse-sPAce cOmplexity in turbulent nearly-reversible plasmas (PAPAO)” (CUP B53C24009610001) funded by the Italian Ministry of University and Research. FC was supported by the ASI project ``Attivita' di Fase A per la missione Plasma Observatory'' (2024-15-HH.0). 
This research was supported by the International Space Science Institute (ISSI) in Bern, through the ISSI International Team project \#24-627
"Impact of stellar wind plasma turbulence on planetary magnetospheres in the heliosphere and beyond".




\newpage
\bibliography{Biblio}{}
\bibliographystyle{aa}





\begin{appendix}

\section{Differential operators}\label{Sec:Diff_OP}
Since the synthetic magnetic field $\mathbf{B}$ is expressed as the curl of a vector potential $\mathbf{A}$, in this section we derive the expressions of differential operators written with respect to curvilinear coordinates. We indicate by $x_i, i=1,2,3$, the Cartesian coordinates of a given point P and by $q^i, i=1,2,3$, the corresponding curvilinear coordinates. In our case, $q^1=\xi$, $q^2=\zeta$ and $q^3=\psi$. We also indicate by $B_i, i=1,2,3$ the Cartesian components of the magnetic field $\mathbf{B}$ (or of any other vector field). The corresponding contravariant $b^i$ and covariant $b_i$ components of $\mathbf{B}$ with respect to the coordinates $q^i$ are related to $B_i$ through the relations:
\begin{equation}\label{contradef}
B_i = b^k \frac{\partial x_i}{\partial q^k} = b^k J_{ik}
\end{equation}
\begin{equation}\label{codef}
b_i = B_k \frac{\partial x_k}{\partial q^i} = B_k J_{ki}\, ,
\end{equation}
where $J_{ik}=\partial x_i/\partial q^k$ are the elements of the Jacobian matrix of the coordinate transformation. Summation over repeated indexes is hereafter understood. It is important to notice that the matrix $J_{ik}$ does not represent a tensor. Therefore, the quantities $J_{ik}$ are not the contravariant components of a tensor. The following relation holds in consequence of the chain rule:
\begin{equation}\label{chain}
\frac{\partial x_i}{\partial q^k} \frac{\partial q^k}{\partial x_n} = \frac{\partial x_i}{\partial x_n} = \delta_{in}\, ,
\end{equation}
which implies
\begin{equation}\label{Jm1}
\left( J^{-1}\right)_{kn} = \frac{\partial q^k}{\partial x_n}\, .
\end{equation}
Moreover, we indicate $\mathcal{J}=\det ( \left\{ J_{ik} \right\})$.

We express the divergence of $\mathbf{B}$ in terms of its contravariant components $b^i$ and of the curvilinear coordinates $q^k$. Using Eq. (\ref{contradef}) and the chain rule we find:
\begin{equation}\label{divB1}
\nabla \cdot \mathbf{B} = \frac{\partial B_k}{\partial x_k} = \frac{\partial}{\partial x_k} \left( b^i \frac{\partial x_k}{\partial q^i} \right) = 
\frac{\partial q^n}{\partial x_k} \frac{\partial}{\partial q^n} \left( b^i \frac{\partial x_k}{\partial q^i} \right)\, ,
\end{equation}
which gives
\begin{equation}\label{divB2}
\nabla \cdot \mathbf{B} = \frac{\partial b^i}{\partial q^n} \frac{\partial q^n}{\partial x_k} \frac{\partial x_k}{\partial q^i} + 
b^i \frac{\partial q^n}{\partial x_k} \frac{\partial}{\partial q^n} \left( \frac{\partial x_k}{\partial q^i} \right)\, .
\end{equation}
We use Eq. (\ref{chain}) in the first term on the RHS of Eq. (\ref{divB2}) and we commute the derivatives $\partial/\partial q^n$ and $\partial/\partial q^i$ in the second term. This gives:
\begin{equation}\label{divB3}
\nabla \cdot \mathbf{B} = \frac{\partial b^i}{\partial q^i} + 
b^i \frac{\partial q^n}{\partial x_k} \frac{\partial}{\partial q^i} \left( \frac{\partial x_k}{\partial q^n} \right)\, .
\end{equation}
Let us consider the quantity multiplying $b^i$ in the last term of Eq. (\ref{divB3}); using the definition of the Jacobian matrix we have:
\begin{equation}\label{secterm1}
\frac{\partial q^n}{\partial x_k} \frac{\partial}{\partial q^i} \left( \frac{\partial x_k}{\partial q^n} \right) =
\left( J^{-1}\right)_{nk} \frac{\partial J_{kn}}{\partial q^i} = 
\left( J^{-1}\right)_{kn} \frac{\partial J_{nk}}{\partial q^i} = 
\frac{1}{\mathcal{J}} {\rm adj} \left( J_{kn}\right) \frac{\partial J_{nk}}{\partial q^i}\, .
\end{equation}
where ${\rm adj} \left( J_{kn}\right) = \mathcal{J} \left( J^{-1}\right)_{kn}$ is the adjugate of the Jacobian matrix. For a given matrix ${\rm M}$ depending on a parameter $t$, the Jacobi formula for the derivative of the determinant holds:
\begin{equation}\label{jacobi}
\frac{d}{dt} \left[ \det {\rm M}(t) \right] = {\rm Tr} \left\{ {\rm adj} \left[ {\rm M}(t)\right] \frac{d {\rm M}(t)}{dt} \right\} = 
{\rm adj} \left[ {\rm M}_{jk} (t)\right] \frac{d {\rm M}_{kj} (t)}{dt}\, .
\end{equation}
Therefore, Eq. (\ref{secterm1}) becomes:
\begin{equation}\label{secterm2}
\frac{\partial q^n}{\partial x_k} \frac{\partial}{\partial q^i} \left( \frac{\partial x_k}{\partial q^n} \right) = 
\frac{1}{\mathcal{J}} \frac{\partial \mathcal{J}}{\partial q^i}\, .
\end{equation}
Inserting the expression (\ref{secterm2}) into Eq. (\ref{divB3}) we obtain:
\begin{equation}\label{divB4}
\nabla \cdot \mathbf{B} = \frac{\partial b^i}{\partial q^i} + 
b^i \frac{1}{\mathcal{J}} \frac{\partial \mathcal{J}}{\partial q^i} =
\frac{1}{\mathcal{J}} \frac{\partial}{\partial q^i} \left( b^i \mathcal{J} \right)\, .
\end{equation}
Based on the above expression, we can set the following form for the contravariant components $b^i$:
\begin{equation}\label{bi}
b^i = \frac{1}{\mathcal{J}} \varepsilon_{ijk} \frac{\partial a_k}{\partial q^j}\, ,
\end{equation}
where $\varepsilon_{ijk}$ is the standard Levi-Civita symbol. Indeed, inserting expression (\ref{bi}) into Eq. (\ref{divB4}) yields
\begin{equation}\label{divver}
\nabla \cdot \mathbf{B} = \frac{1}{\mathcal{J}} \varepsilon_{ijk} \frac{\partial^2 a_k}{\partial q^i \partial q^j} = 0\, .
\end{equation}
Therefore, the expression (\ref{bi}) gives a solenoidal field, for any choice of the functions $a_k=a_k(q^1,q^2,q^3)$. Using  expression (\ref{bi}) in Eq. (\ref{contradef}), we finally obtain the form for the magnetic field Cartesian components:
\begin{equation}\label{cartB}
B_i = \frac{J_{ik}}{\mathcal{J}} \varepsilon_{klm} \frac{\partial a_m}{\partial q^l}\, .
\end{equation}

\subsection{Specific forms}

In our case, the transformation between Cartesian coordinates $(x,y,z)=(x_1,x_2,x_3)$ and curvilinear coordinates $(q^1,q^2,q^3)=(\xi,\zeta,\psi)$ is done by an intermediate step involving spherical coordinates $(s_1,s_2,s_3)=(\rho,\theta,\varphi)$. The latter are related to Cartesian coordinates by
\begin{eqnarray}\label{cartvsspher}
   x_1 &=& \rho \sin \theta \cos \varphi \nonumber \\
   x_2 &=& \rho \sin \theta \sin \varphi \\
   x_3 &=& \rho \cos \theta \nonumber\, ,
\end{eqnarray}
while spherical coordinates are related to curvilinear coordinates by the relations (\ref{rhovsxi}), (\ref{thetavsxizeta}) and (\ref{phivsxizetapsi}). These can be re-written in the following form:
\begin{equation}\label{rhovsxi2}
\rho = \frac{\rho_0}{(1-\alpha_0 \xi)^{1/(h-1)}}
\end{equation}
\begin{equation}\label{thetavsxizeta2}
\theta = \frac{\pi}{2} - \frac{\beta_0 \zeta}{1-\alpha_0 \xi}
\end{equation}
\begin{equation}\label{phivsxizetapsi2}
\varphi = \frac{k \rho_0}{(1-\alpha_0 \xi)^{1/h-1}} + 
\frac{\beta_0 \psi}{(1-\alpha_0 \xi) \cos \left( \displaystyle{\frac{\beta_0 \zeta}{1-\alpha_0 \xi}}\right)}
\end{equation}
where, to simplify the expressions, we introduced the constant quantities $\alpha_0=1-\rho_0^{h-1}$ and $\beta_0=(1-\rho_0^{h-1})/(h-1)=\alpha_0/(h-1)$.

Using the chain rule, the Jacobian matrix can be written as a product of two matrices:
\begin{equation}\label{JRD}
J_{ik}=\frac{\partial x_i}{\partial q^k} = \frac{\partial x_i}{\partial s_n} \frac{\partial s_n}{\partial q^k} = R_{in} S_{nk}\, .
\end{equation}
where $R_{in}=\partial x_i/\partial s_n$ and $S_{nk}=\partial s_n/\partial q^k$ are the two intermediate Jacobian matrices. The determinant of $J_{ik}$ is given by $\mathcal{J}=\mathcal{R}\mathcal{S}$, with $\mathcal{R}=\det(R_{in})$ and $\mathcal{S}=\det(S_{nk})$. The specific form of the Jacobian matrix $R_{in}$ is derived from Eqs. (\ref{cartvsspher}):
\begin{equation}\label{Rin}
R_{in} = 
\begin{bmatrix}
\sin \theta \cos \varphi & \rho \cos \theta \cos \varphi & -\rho \sin \theta \sin \varphi \\
\sin \theta \sin \varphi & \rho \cos \theta \sin \varphi & \rho \sin \theta \cos \varphi \\
\cos \theta & -\rho \sin \theta & 0\, ,
\end{bmatrix}
\end{equation}
while $\mathcal{R}=\rho^2 \sin \theta$. The elements of the Jacobian matrix $S_{nk}$ are obtained from Eqs. (\ref{rhovsxi2})-(\ref{phivsxizetapsi2}) and are given by:
\begin{eqnarray}\label{Srow1}
S_{11}=\frac{\partial \rho}{\partial \xi} &=& \frac{\rho_0 \alpha_0}{(h-1)(1-\alpha_0 \xi)^{\frac{h}{h-1}}} \;\; , \nonumber \\
S_{12} &=& \frac{\partial \rho}{\partial \zeta} = 0 \;\;\; , \;\;\; S_{13}=\frac{\partial \rho}{\partial \psi} =0
\end{eqnarray}
\begin{eqnarray}\label{Srow2}
S_{21}=\frac{\partial \theta}{\partial \xi} &=& -\frac{\alpha_0 \beta_0 \zeta}{(1-\alpha_0 \xi)^2} \;\;\; , \;\;\; 
S_{22}=\frac{\partial \theta}{\partial \zeta} = -\frac{\beta_0}{1-\alpha_0 \xi} \;\; , \nonumber \\
S_{23} &=& \frac{\partial \theta}{\partial \psi} = 0
\end{eqnarray}
\begin{eqnarray}\label{S31}
S_{31}=\frac{\partial \varphi}{\partial \xi}
= \frac{k \rho_0 \alpha_0}{(h-1)(1-\alpha_0 \xi)^{\frac{h}{h-1}}} + \frac{\beta_0 \alpha_0 \psi}{(1-\alpha_0\xi)^2 \cos \left( \displaystyle{\frac{\beta_0 \zeta}{1-\alpha_0 \xi}}\right)} \nonumber \\
+ \frac{\beta_0^2 \alpha_0 \zeta \psi \sin \left( \displaystyle{\frac{\beta_0 \zeta}{1-\alpha_0 \xi}}\right)}{(1-\alpha_0 \xi)^3 \cos^2 \left( \displaystyle{\frac{\beta_0 \zeta}{1-\alpha_0 \xi}}\right)}
\end{eqnarray}
\begin{equation}\label{S32}
S_{32} = \frac{\partial \varphi}{\partial \zeta} = \frac{\beta_0^2 \psi \sin \left( \displaystyle{\frac{\beta_0 \zeta}{1-\alpha_0 \xi}}\right)}{(1-\alpha_0\xi)^2 \cos^2 \left( \displaystyle{\frac{\beta_0 \zeta}{1-\alpha_0 \xi}}\right)}
\end{equation}
\begin{equation}\label{S33}
S_{33} = \frac{\beta_0}{(1-\alpha_0 \xi) \cos \left( \displaystyle{\frac{\beta_0 \zeta}{1-\alpha_0 \xi}}\right)}\, .
\end{equation}
Finally, since $S_{nk}$ is a triangular matrix, its determinant is given by $\mathcal{S}=S_{11}S_{22}S_{33}$:
\begin{equation}\label{S}
\mathcal{S}= -\frac{\alpha_0 \beta_0^2 \rho_0}{(h-1) (1-\alpha_0\xi)^{\frac{3h-2}{h-1}} \cos \left( \displaystyle{\frac{\beta_0 \zeta}{1-\alpha_0 \xi}}\right)}
\end{equation}

\end{appendix}

\end{document}